\documentclass[a4paper,11pt]{article}
\pdfoutput=1 % if your are submitting a pdflatex (i.e. if you have
\usepackage{jheppub} % for details on the use of the package, please
\usepackage[T1]{fontenc}
\usepackage[italian,english]{babel}
\usepackage{hyperref}
\usepackage{ifpdf}
\usepackage{subfigure}
\usepackage{amssymb}
\usepackage{amsfonts}
\usepackage{epsf}
\usepackage{rotating}
\usepackage{graphicx}
\usepackage{amsmath}
\usepackage{fancyhdr}
\usepackage{lineno}
\usepackage{babel}
\usepackage{graphics}
\usepackage{pstricks}
\usepackage{color}
\usepackage{multirow}
\usepackage{url}
\usepackage{alphabeta}
\newcommand{\nn}{\nonumber}
\newcommand{\lsim}{\mathrel{\mathop{\kern 0pt \rlap
  {\raise.2ex\hbox{$<$}}}
  \lower.9ex\hbox{\kern-.190em $\sim$}}}
\newcommand{\gsim}{\mathrel{\mathop{\kern 0pt \rlap
  {\raise.2ex\hbox{$>$}}}
  \lower.9ex\hbox{\kern-.190em $\sim$}}}

\newcommand{\be}{\begin{equation}}
\newcommand{\ee}{\end{equation}}
\newcommand{\bea}{\begin{eqnarray}}
\newcommand{\eea}{\end{eqnarray}}

\def\ptmiss{\not\!\!{p_T}}
\title{\boldmath Phenomenology of Higgs bosons in inverse seesaw model with Type-X two Higgs doublet at the LHC}

\author[a]{Priyotosh Bandyopadhyay}

\author[b]{Eung Jin Chun}
\author[c]{Rusa Mandal}
\affiliation[a]{Indian Institute of Technology Hyderabad, Kandi,  Sangareddy-502287, Telengana, India}

\affiliation[b]{Korea Institute for Advanced Study, Seoul 130-722, Korea}
\affiliation[c]{IFIC, Universitat de Val$\grave{e}$ncia-CSIC, Apt. Correus 22085, E-46071 Val$\grave{e}$ncia, Spain}
\emailAdd{bpriyo@iith.ac.in}
\emailAdd{ejchun@kias.re.kr}
\emailAdd{Rusa.Mandal@ific.uv.es}
\preprint{IITH-PH-0001/19}
\abstract{ 
Type-X two Higgs doublet model is known to explain the muon $g-2$ anomaly with a relatively light charged Higgs boson at large $\tan\beta$. The light charged Higgs boson has been searched in the main $\tau \nu$ mode at the colliders. Invoking a scenario of inverse seesaw as the origin of neutrino masses and mixing, the charged Higgs boson can decay additionally to right-handed neutrinos which leads to interesting phenomenology.  Considering generic lepton flavour violating signatures at the final states,  a $5\sigma$ discovery can be achieved with the early data of LHC, at 14 TeV, for relatively large inverse seesaw Yukawa coupling $Y_N$. The very light pseudoscalar  and charged Higgs boson  mass  reconstruction are performed using the new modes and the results look promising. The inverse seesaw Yukawa coupling is shown to be probed down to $Y_N \sim 0.2$ at HL LHC with 3000 fb$^{-1}$.
}

\begin{document}
\maketitle
\flushbottom

\section{Introduction}
Non-observation of flavour changing neutral currents classifies Two-Higgs-Doublet Models (2HDMs) to four different categories which differ by the pattern of Higgs doublets' interaction to fermions~\cite{Gunion:2002zf}. A discrete symmetry $Z_2$ is imposed on these models under which the Higgs doublets and fermions carry different parities. The well-known nomenclature is ``Type-I", ``Type-II", ``lepton-specific"(or ``Type-X") and ``flipped"(or ``Type-Y")  2HDMs. An interesting scenario is the Type-X 2HDM which can explain the $g-2$ anomaly \cite{Broggio:2014mna} by evading all the collider bounds for high $\tan{\beta}$ regime \cite{Jinsu,Cao:2009as}. An extension of such scenario with a scalar dark matter candidate also provides interesting signature in indirect measurements~\cite{Bandyopadhyay:2017tlq}.
In this large $\tan{\beta}$ limit, due to the suppression in couplings of the heavy Higgs bosons to quarks (which affects their production cross section at the LHC), the popularity of this model is depreciated from collider searches point of view. An alleviation is possible in presence of a light pseudoscalar which opens the decay modes to  $A, Z$ and $A, W^\pm$ for the heavier Higgs bosons, $H$ and $H^\pm$, respectively. The decay width of $H^\pm \to A W^\pm$ is independent of $\tan{\beta}$ and depends only on the gauge coupling, thus the branching fraction in this mode becomes very prominent at high $\tan{\beta}$ region where the other decay modes are suppressed. In the context of Type-X, the parameter space with a light pseudoscalar boson and larger values of $\tan{\beta}$ has been investigated in various direct and indirect searches \cite{Cao:2009as}.  This decay mode of charged Higgs has also  been looked into for Type-I and Type-II 2HDM scenarios~\cite{other}.

The presence of light pseudoscalar is very natural in $Z_3$ symmetric superpotential viz., NMSSM \cite {NMSSM, NMSSMCH} and  Triplet-Singlet-extended MSSM \cite{TNSSM}, where it arises as pseudo-Nambu-Goldstone mode and the studies for the decay of charged Higgs to this light pseudoscalar are performed as well.  The muti-lepton and multi-tau final states are also investigated in the context of triplet-like charged Higgs bosons with the bounds from $B\to X_s \gamma$ \cite{mtau}.  However, such studies do not have the right-handed neutrino (RHN) in the final states. Construction of the RHN thus becomes very crucial in order to distinguish our scenario from the rest. As an additional benefit, non-democratic lepton-flavour signature arise at the final states which is a smoking gun signature of our model. 

The signature arising from the RHN can be enhanced at the colliders with a relatively larger Yukawa coupling of a RHN via inverse seesaw mechanism. This also  enriches the phenomenology and the search for such Type-X charged Higgs boson. In an inverse seesaw framework the RHN is a pseudo-Dirac fermion allowing an $\mathcal{O}(1)$ coupling with the Higgs bosons. This enables us to search for the decays of the charged Higgs boson into charged lepton and RHN, where the RHN can further decay into the following modes: charged  lepton/neutrino and gauge boson, neutrino and Higgs boson, as well as light pseudoscalar and neutrino.  The right-handed neutrinos decays to charged Higgs can be seen in the context of other scenarios \cite{ty2, uNs} but can only be enhance in inverse-seesaw case due to relatively large Yukawa coupling
In this article we are mostly interested in probing the decay modes with charged lepton, gauge boson and also the pseudoscalar, which is generic in Type X 2HDM, at the LHC.

The paper is organized as follows. In Sec. \ref{sec:model} we briefly describe the model. By studying the parameter space  allowed by several measurements, we chose the benchmark points in Sec. \ref{sec:BP}.  The collider phenomenology is discussed in Sec. \ref{sec:simulation} and the corresponding results are presented in Sec.~\ref{sec:results} including a discussion with the phenomenology of light pseudoscalar which is different compared to the other benchmark cases. In Sec. \ref{sec:Chiggs} we discuss the charged Higgs mass reconstruction and the reach at current and future LHC.  Finally in Sec.~\ref{sec:conclusion} we present the concluding remarks.

\section{The Model}
\label{sec:model}
 We consider three generations of $N_R$  and $S_2$, the two Majorana neutrinos forming a pseudo-Dirac fermion, which are singlet under the SM gauge group. Here $N_R$ couples to the left-handed active neutrino $\nu_L$ via Yukawa coupling $Y_N$ shown in Eq.~\ref{lag}, which can be $\mathcal{O}(1)$ in the inverse seesaw mechanism \cite{iss, bliss, pbiss}.  The other Majorana neutrino $S_2$ does not have any direct coupling to the SM sectors but mixes with $N_R$ via a mass mixing term proportional to $M_N$ (the fifth term in Eq.~\ref{lag}). It has a Majorana mass term $\mu$ which can be very small motived from the breaking of higher $U(1)_{B-L}$ gauge group \cite{bliss}. 
 
 Here we invoke the inverse seesaw mechanism in the Type-X 2HDM, which is capable in explaining the muon $g-2$ anomaly at $2\sigma$ level in presence of a light pseudoscalar \cite{Broggio:2014mna}. In this case the charged Higgs boson can also be very light unlike Type-II 2HDM, which suffers from indirect bounds arising from $B \to X_s \gamma$ \cite{bsg} mode.  In Eq.~\ref{lag} we see that the leptons interact to the Higgs doublet $\Phi_1$ whereas the quarks couple to $\Phi_2$. Interestingly, the RHN $N_R$ can couple to both $\Phi_1$ and $\Phi_2$ and we call such extensions as Type-X and Type-X$^\prime$, respectively. In the succeeding sections we focus on Type-X parameter space for collider phenomenology.
 \bea\label{lag}
 -\mathcal{L}&=& (Y_u \bar{Q}_L\tilde{\Phi}_2 u_R \, +\, Y_d \bar{Q}_L\Phi_{2} d_R \, + Y_l \bar{\ell}_L\Phi_{1} e_R \, + Y_{N}^{(\prime)} \bar{\ell}_L\tilde{\Phi}_{1,2}N_R\, \nonumber \\
  &&+ \, M_N \bar{N}^c_R S_2\, + {\rm h.c.})+\, \mu \bar{S}^c_2S_2 + V(\Phi_1, \Phi_2). \, 
 \eea
 Note that $Y_N^{(\prime)} $ corresponds to $3\times 3$ Yukawa matrix which couples the RHNs to different SM lepton generations.  The Higgs doublets $ \Phi_{1,2}$ are given by Eq.~\ref{hd} and $\tilde{\Phi}_2=i\sigma_2\phi_2^*$ where $\sigma_2$ is the Pauli matrix. 
 \begin{equation}\label{hd}
 \Phi_{1,2} =\begin{pmatrix} 
 \phi^+_{1,2}  \\
 {\frac{1}{\sqrt{2}}}\left(v_{1,2} + h_{1,2} + i a_{1,2}\right)
 \end{pmatrix}.
 \end{equation}
The neutrino mass terms in the Lagrangian can be written as
 \be\label{lmass}
 -\mathcal{L}^\nu_{m}= \mu \bar{S}^c_2S_2 + m_D \bar{\nu}_L N_R + M_N \bar{N}^c_R S_2 + {\rm h.c.},
 \ee
 where $m_D= Y_N^{(\prime)} \, v_{1,2}/\sqrt{2}$ for Type-X and Type-X$^\prime$, respectively. In the basis of ${\nu^c_L, N_R, S_2}$, the $9 \times 9$ neutrino mass matrix takes the form as
 \be\label{massm}
 m_{\nu}=\begin{pmatrix}
 0 \, \quad m_D\, \quad 0 \\
 m^T_D\, \quad 0\, \quad M_N\\
 0\,\quad   M^T_N\, \quad  \mu  
 \end{pmatrix}.
 \ee
 Diagonalizing the matrix (Eq.~\ref{massm}) we have three categories for neutrinos where the masses are given by 
 \bea\label{neueigen}
 m_{\nu_{\ell}}=m_D M^{-1}_N\mu (M^T_N)m^T_D,\\
 m^2_{N_H}=m^2_{N_{H'}}= M^2_N +m^2_D.
 \eea	
We designate these nearly mass degenerate Majorana eigenstates as $N_i$, where  $i\in\{1-6\}$, for the rest of the paper.

Having two Higgs doublets $\Phi_{1,2}$, we write the $Z_2$ symmetric scalar potential as
\begin{eqnarray} \label{scalar-potential}
V &=& m_{11}^2 |\Phi_1|^2
+ m_{22}^2 |\Phi_2|^2 - m_{12}^2 (\Phi_1^\dagger \Phi_2 + \Phi_1 \Phi_2^\dagger) \nonumber \\
& +& {\lambda_1\over2} |\Phi_1|^4 + {\lambda_2 \over 2} |\Phi_2|^4
+ \lambda_3 |\Phi_1|^2 |\Phi_2|^2 + \lambda_4 |\Phi_1^\dagger \Phi_2|^2 + {\lambda_5 \over 2}
\left[ (\Phi_1^\dagger \Phi_2)^2 + (\Phi_1 \Phi_2^\dagger)^2\right],
\end{eqnarray}
where a (soft) $Z_2$ breaking term $m^2_{12}$ is introduced.
Minimization of the scalar potential determines the vacuum expectation values $\langle \Phi^0_{1,2} \rangle = v_{1,2}/\sqrt{2}$ around which the Higgs doublet fields are expanded. The model contains five physical fields denoted by $H^\pm,\, A,\, H$ and $h$ in the mass basis and their orthogonal combinations are the corresponding Goldstone modes $G^{\pm, 0}$. The mass basis and gauge basis are related by the following rotation matrices
\bea\label{basis}
\begin{pmatrix} 
	H^\pm \\
	G^\pm
\end{pmatrix}&=&\begin{pmatrix} 
\sin{\beta} & \cos{\beta}\\
-\cos{\beta}& \sin{\beta} 
\end{pmatrix}\begin{pmatrix} 
H^\pm_1 \\
H^\pm_2
\end{pmatrix}, \quad \quad \begin{pmatrix} 
A\\
G^0
\end{pmatrix}=\begin{pmatrix} 
\sin{\beta} & \cos{\beta}\\
-\cos{\beta}& \sin{\beta} 
\end{pmatrix}\begin{pmatrix} 
a_1 \\
a_2
\end{pmatrix},\nn \\
 \begin{pmatrix} 
	H\\
	h
\end{pmatrix}&=&\begin{pmatrix} 
	\cos{\alpha} & \sin{\alpha}\\
	-\sin{\alpha}& \cos{\alpha} 
\end{pmatrix}\begin{pmatrix} 
	h_1 \\
	h_2
\end{pmatrix},
\eea
where the angle $\beta$ is defined as $t_\beta\equiv \tan\beta =v_2/v_1$.
The neutral CP-even Higgs bosons are diagonalized such that $h\, (H)$ denotes the lighter (heavier) state.

The gauge interaction of the Higgs bosons $h$ and $H$ are given  by
${\cal L}_{\rm gauge} \simeq g_V m_V \big(s_{\beta -  \alpha} h + c_{\beta-\alpha} H \big) VV$
where $V=W^\pm$ or $Z$. In the case of $h$ being 125 GeV Higgs boson, the SM limit corresponds to $s_{\beta-\alpha} \to 1$.  Indeed, LHC finds  $c_{\beta-\alpha} \ll 1 $ in all the 2HDMs confirming the SM-like property of the 125 GeV boson \cite{CMS:2015kwa}.

Normalizing the Yukawa couplings of the neutral bosons and a fermion $f$  by $m_f/v$ factor where $v=\sqrt{v_1^2+v_2^2} = 246$ GeV, we obtain the following couplings of the respective Yukawa terms.
\bea\label{yuk}
\begin{aligned}
\bar{q_L} H^\pm q{'}_R&: -i \frac{\sqrt{2}\cot{\beta}}{v}\big[-m_u \bar{d_L}H^- u_R \, +\, m_d \bar{u_L}H^+d_R \, +\, {\rm h.c.}\big],\\
\bar{q_L} A q_R&:-\frac{\cot{\beta}}{v} \big[- m_u\bar{u_L} A u_R  \, +\, m_d \bar{d_L}Ad_R + {\rm h.c.}\big],\\
\bar{q_L} h q_R&:-\frac{-i \cos{\alpha}}{v\sin{\beta}} \big[m_u \bar{u_L} h u_R  \, +\, m_d \bar{d_L}hd_R + {\rm h.c.}\big],\\
\bar{q_L} H q_R&:-\frac{-i \sin{\alpha}}{v\sin{\beta}} \big[m_u \bar{u_L} H u_R  \, +\, m_d \bar{d_L}H d_R + {\rm h.c.}\big], \\
\bar{\ell}_L H \ell_R&:\frac{-i m_{\ell}\cos{\alpha}}{v\cos{\beta}} \big[ \bar{\ell}_L H \ell_R  \,  + \,{\rm h.c.}\big],\\
\bar{\ell}_L h \ell_R&:\frac{i m_{\ell}\sin{\alpha}}{v\cos{\beta}}  \big[\bar{\ell}_L H \ell_R  \,  + \, {\rm h.c.}\big],\\
\bar{\ell}_L A \ell_R&:\frac{m_{\ell}\tan{\beta}}{v} \big[ \bar{\ell}_L A \ell_R  \,  + \,{\rm h.c.}\big],\\
\bar{\nu}H^+ \ell_R&:\frac{-i m_{\ell}\tan{\beta}}{v} \big[ \bar{\nu} H^+ \ell_R  \, + \,{\rm h.c.}\big].
\end{aligned}
\eea

However, as we are interested in Type-X 2HDM, the choice of interaction term of the RHN with the Higgs doublets is very crucial. For that reason we consider two cases as mentioned before and is explicitly shown in Eq.~\ref{lag2},  where we name it Type-X extension when the RHN couples to $\tilde{\Phi}_1$, like the SM leptons, and Type-X$^\prime$ when it couples to $\tilde{\Phi}_2$. 
\be\label{lag2}
-\mathcal{L}^{\text{Type-X}}_{int}=  Y_{N} \bar{\ell}_L\tilde{\Phi}_{1}N_R \quad \text{and}\quad 
-\mathcal{L}^{\text{Type-X$^\prime$}}_{int}=  Y_{N}^\prime \bar{\ell}_L\tilde{\Phi}_{2}N_R.
\ee
Depending on the Type-X or Type-X$^\prime$ extension, the decays of RHN will have very different behavior with $\tan{\beta}$ variation. Below we list the relevant couplings of RHN with the other fields present in the model where the set in Eq.~\ref{yuk2} is for the Type-X case and Eq.~\ref{yuk3} refers to Type-X$^\prime$ extension.
\bea\label{yuk2}
\begin{aligned}
\bar{\ell}_L H^- N_R&:&i Y_N\sin{\beta}\big[  \bar{\ell}_L H^- N_R  \, + \,{\rm h.c.} \big], \\
\bar{\nu}_L h N_R&:&\frac{i Y_N \sin{\alpha}}{\sqrt{2}} \big[ \bar{\nu}_L h N_R  \, + \,{\rm h.c.} \big], \\
\bar{\nu}_L H N_R&:&\frac{-i Y_N \cos{\alpha}}{\sqrt{2}} \big[ \bar{\nu}_L H N_R  \, + \,{\rm h.c.} \big], \\
\bar{\nu}_L A N_R&:&\frac{-Y_N \sin{\beta}}{\sqrt{2}} \big[ \bar{\nu}_L A N_R  \, + \,{\rm h.c.} \big]. 
\end{aligned}
\eea
\bea\label{yuk3}
\begin{aligned}
\bar{\ell}_L H^- N_R&:&i Y_N^\prime\cos{\beta}\big[  \bar{\ell}_L H^- N_R  \, + \,{\rm h.c.} \big], \\
\bar{\nu}_L h N_R&:&\frac{-i Y_N^\prime \cos{\alpha}}{\sqrt{2}} \big[ \bar{\nu}_L h N_R  \, + \,{\rm h.c.} \big], \\
\bar{\nu}_L H N_R&:&\frac{-i Y_N^\prime \sin{\alpha}}{\sqrt{2}} \big[ \bar{\nu}_L H N_R  \, + \,{\rm h.c.} \big], \\
\bar{\nu}_L A N_R&:&\frac{-Y_N^\prime \cos{\beta}}{\sqrt{2}} \big[ \bar{\nu}_L A N_R  \, + \,{\rm h.c.} \big].
\end{aligned}
\eea
It can be seen that in high $\tan{\beta}$ region, the decay modes $H^\pm \to \ell_L N_R$ and  $N_R \to A \nu_L$, which are of our special interests, are enhanced only in Type-X extension and thus we examine the Type-X extension with RHNs in the rest of the paper. We also note that the decay $H^\pm \to A W^\pm$ is governed only by the weak gauge coupling $g_2 $ in all 2HDM scenarios, however, due to the dependency of $H^\pm \to \ell_L N_R$ on $\tan{\beta}$ values, the partial branching fraction for $H^\pm \to A W^\pm$ may vary which has important consequences in collider studies explored in the subsequent sections.

\section{Benchmark points}
\label{sec:BP}
To probe the exotic decays of the other (apart from the SM like one) Higgs bosons, specially the charged Higgs boson we choose some  benchmark points for collider study. The $\mu \to e \gamma$ bounds from MEG collaboration \cite{MEG} can be avoided by choosing the diagonal Yukawa for the RHNs.  The EWPT also is allowed in the alignment limit \cite{Erler, Blas, Aguila}. In principle for collider searches we can choose the Yukawa responsible for inverse seesaw, $Y_{N_i}$ of $\mathcal{O}(1)$. For the current study we choose $Y_{N_i}=0.5$ for the democratic benchmark points viz. BP1, BP2 and BP3. However, for BP4 we choose $Y_{N_1}=0.5,\, Y_{N_{2,3}}=0.1$. In Table~\ref{bps} we present the mass spectra  and other relevant parameters for these different benchmark points for the collider study. Amidst of such points BP3 has a light pseudoscalar with $m_A \sim 50$\,GeV.
 %%%%%%%%%%%%%%%%% Benchmark points%%%%%%%%%%%%%%%%%%%%
\begin{table}[!h]
	\begin{center}
		\renewcommand{\arraystretch}{1.5}
		\begin{tabular}{||c||c|c|c|c||}
			\hline\hline
			Benchmark&BP1&BP2&BP3&BP4\\
			Points & &&&\\ \hline\hline
			$m_{h}$ &125.5 &125.5&125.5&125.5 \\
			\hline
			$m_{H}$ &250.1  &250.1 & 250.1& 250.1\\
			\hline
			$m_{A}$ & 100.0& 200.1& 49.6 & 100.0\\
			\hline
			$m_{H^\pm}$ &250.1 & 250.1&250.1& 250.1\\
			\hline
			$m_N$ &98.0& 100.0& 100.0&100.0 \\
			\hline
		  $Y_{N_1}$ &0.5&0.5&0.5&0.5\\
			\hline
			$Y_{N_{2,3}}$ &0.5&0.5&0.5&0.1\\
			\hline
			$\tan{\beta}$ &50.0&50.0&50.0&50.0\\
			\hline
			\hline
		\end{tabular}
		\caption{Benchmark points for a collider study consistent with $m_h\sim 125$ GeV the SM like Higgs mass and with the $2\sigma$ limits of $h\to WW^*, ZZ^*, \gamma\gamma$ \cite{CMS:2015kwa}. Here we have only considered the non-zero diagonal Yukawa couplings i.e., $Y_{N_{1,2,3}}\equiv Y_{N_{11,22, 33}}$, respectively. }\label{bps}
	\end{center}
\end{table}

%%%%%%%%%%%%  Decay Branching fraction of H+ %%%%%%%%%%%%%%%%%%%%
\begin{table}
	\begin{center}
		\renewcommand{\arraystretch}{1.5}
		\begin{tabular}{||c||c|c|c|c||}
			\hline\hline
			Benchmark&BP1&BP2&BP3&BP4\\
			Points & &&&\\ \hline\hline
			$A W^\pm$ & 0.30 &0.00 &0.44&0.42 \\
			\hline
			$\tau\, \nu_\tau$ &0.22 & 0.34&0.17&0.33\\
			\hline
			$e^\pm N$ &0.16&0.22&0.13&0.23\\
			\hline
			$\mu^\pm N$ &0.16&0.22&0.13&0.01\\
			\hline
			$\tau^\pm N$ &0.16&0.22&0.13&0.01\\
			\hline
			\hline
		\end{tabular}
		\caption{Branching fraction for $H^\pm$ for collider study at the LHC for $Y_N=0.5\,(Y_{N_1}=0.5, Y_{N_{2,3}}=0.1)$ for BP1 - BP3 (BP4). Here  $N$ corresponds to inclusive of heavy neutrinos, i.e.,  $\sum\limits_i N_i$.} \label{brch}
	\end{center}
\end{table}
%%%%%%%%%%%%  

%%%%%%%%%%PSEUDOSCALAR A BRANING%%%%%%%%
\begin{table}
	\begin{center}
		\renewcommand{\arraystretch}{1.5}
		\begin{tabular}{||c||c|c|c|c||}
			\hline\hline
			Benchmark&BP1&BP2&BP3&BP4\\
			Points & &&&\\ \hline\hline
			$\tau\bar{\tau}$ & 0.99& 0.38&0.99&0.99 \\
			\hline
			%			$b\bar{b}$ &0.00&0.00& -&0.00\\
			\hline
			$\sum\limits_i N_i\nu_i$  & 0.01 &0.62 &0.00&$\sim10^{-3}$\\
			\hline
			\hline
		\end{tabular}
		\caption{Branching fraction for $A$ for the benchmark points for collider study at the LHC.}\label{bra1}
	\end{center}
\end{table}
%%%%%%%%%%%%  

\subsection{Decay branching fractions}
As discussed in the introduction the light charged Higgs boson $<500$ Gev is still allowed for Type-X compared to Type-II 2HDM. For the given BPs, we have chosen a charged Higgs boson with mass of 250 GeV, which opens up a large parameter space explaing the muon $g-2$ deviation \cite{Jinsu}. The pseudoscalar mass varies from 49.6 GeV to 200 GeV depending on the benchmark points.  
Table~\ref{brch} present the decay branching fractions for the charged Higgs bosons for the benchmark points. For all benchmark points except for BP2, we see that $A W^\pm$ is the dominant mode as for large $\tan{\beta}$ the $t\,b$ mode is suppressed which can be seen from Eq.~\eqref{yuk}.  Apart from $AW^\pm$ modes, the decay of charged Higgs boson to RHN and charged lepton can also be significant. For BP4, due of the choice of non-democratic Yukawa couplings i.e., $Y_{N_1}=0.5,\, Y_{N_{2,3}}=0.1$, the charged Higgs dominantly decays  only to $N_1 e^\pm$ states.

The light pseudoscalar mostly decays to tau anti-tau pair as shown in Table.~\ref{bra1}. The $b\bar{b}$ mode is suppressed due to large value of $\tan{\beta}= 50$ for all four benchmark points.  However, as for BP2 $m_A=200$ GeV, the branching fraction to $N \nu$ is $62\%$ due to the available phase space compared to other BPs. For BP3, this mode is not kinematically allowed. 

Finally we notice that the branching ratios for $H$ also changes compared to  the 2HDM case as $N_i \nu_i$ modes are now open and have substantial branching fraction in this channel which can be read from Table~\ref{brh2}. Due to the significant reduction in decay branching to $ZZ$ and $W^\pm W^\mp$ final states, which are actually vanishing in this case, the heavy Higgs boson can easily evade the current bounds for various experimental searches \cite{lhcheavyHiggs}.

The RHNs in this case mostly decay to $W^\pm \ell^\mp$ and the corresponding branching fraction is given in Table~\ref{brN}. The decays to final states with Higgs bosons are kinematically disallowed for all BPs and in the case of BP3, the RHNs decay completely to the light pseudoscalar and neutrino channel.

%%%%%%%%%%PSEUDOSCALAR h2 BRANING%%%%%%%%
\begin{table}
	\begin{center}
		\renewcommand{\arraystretch}{1.5}
		\begin{tabular}{||c||c|c|c|c||}
			\hline\hline
			Benchmark&BP1&BP2&BP3&BP4\\
			Points & &&&\\ \hline\hline
			$A Z$ & 0.26&0.00  &0.41&0.38\\
			\hline
			%			$A A$ & 0.00 &0.00 & $4.8\times 10^{-4}$&0.00 \\
			%			\hline
			$\tau\bar{\tau}$ & 0.24&0.33 &0.19&0.36 \\
			\hline
			%			$\mu\bar{\mu}$ & 0.00 & 0.00&$6.7\times 10^{-4}$& 0.00\\
			%			\hline
			%			$b\bar{b}$ &0.00&0.00&$3.1\times 10^{-8}$&0.00 \\
			%			\hline
			$\sum\limits_i N_i\nu_i$  & 0.50&0.67&0.39&0.26\\
			\hline
			%			$ZZ$  &0.00  &0.00 &$1.3\times 10^{-4}$&0.00\\
			%			\hline
			%			$W^\pm W^\mp$  &0.00  &0.00& $3.4\times 10^{-4}$&0.00\\
			\hline
		\end{tabular}
		\caption{Branching fraction for $H$ for the benchmark points for collider study at the LHC.}\label{brh2}
	\end{center}
\end{table}
%%%%%%%%%%%%  

%%%%%%%%%%RHN  BRANING%%%%%%%%
\begin{table}
	\begin{center}
		\renewcommand{\arraystretch}{1.5}
		\begin{tabular}{||c||c|c|c|c||}
			\hline\hline
			Benchmark&BP1&BP2&BP3&BP4\\
			Points & &&&\\ \hline\hline
			$W^\pm \ell^\mp$ & 0.91&0.88 &&0.91 \\
			\hline
%				$W^\pm e^\mp$ & 0.91&0.88 &&0.91 \\
%			\hline
%				$W^\pm \mu^\mp$ & 0.91&0.88 &&0.91 \\
%			\hline
%				$W^\pm \tau^\mp$ & 0.91&0.88 &&0.91 \\
			\hline
			$Z \nu$ &0.09 &0.12& &0.09\\
			\hline
%			$h \nu$  & 0.00&0.00 &&0.00\\
%				\hline
			$A \nu$  & 0.00&0.00 &1.00&0.00\\
			\hline
			\hline
		\end{tabular}
		\caption{Branching fraction for $N_i$ for the benchmark points for collider study at the LHC. Here $ \ell^\mp$ spans over all three charged leptons namely, $e, \mu , \tau$ depending on the choice of $N_i$.  }\label{brN}
	\end{center}
\end{table}
%%%%%%%%%%%%  

\subsection{Cross-section}
The model considered in this paper is implemented in SARAH \cite{sarah} where the corresponding files for CalcHEP \cite{calchep} are generated. The cross-sections for the Higgs bosons are calculated using CalcHEP with $\sqrt{\hat{s}}$ and CTEQ6L \cite{cteq} are chosen as the renormalization and factorization scale and PDF, respectively. The largest cross-sections arise for $AH$ and $AH^\pm$ modes. The production cross-sections for BP1 and BP4 are the same as the mass spectrum and the Higgs couplings are the same.  Below we discuss the final state topologies that can be probed at the LHC for the chosen benchmark points.

%%%%%%%%%%cross-section %%%%%%%%
\begin{table}
	\begin{center}
		\renewcommand{\arraystretch}{1.5}
		\begin{tabular}{||c||c|c|c|c||}
			\hline\hline
			Benchmark&BP1&BP2&BP3&BP4\\
			Points & &&&\\ \hline\hline
			$A H$ & 26.8 &11.5& 39.5&26.8\\
			\hline
			$A H^\pm$ &49.7 & 21.8 &72.8&49.7  \\
			\hline
			$H H^\pm$ & 14.7 &14.7 & 14.7 &14.7 \\
			\hline
			$H^\pm H^\mp$ & 8.1 &8.1 &8.1&8.1  \\
			\hline
%			$W H^\pm$ &0.1&& &0.1\\
%			\hline
%			$Z H$  & $10^{-9}$ & &&$10^{-9}$\\
%			\hline
			\hline
		\end{tabular}
		\caption{Tree-level cross-section for the benchmark points obtained by CalcHEP \cite{calchep} in the units of fb  at the LHC with center of mass energy of 14\,TeV, $\sqrt{\hat{s}}$ as renormalization and factorization scale, and CTEQ6L \cite{cteq} as PDF.}\label{cross}
	\end{center}
\end{table}
%%%%%%%%%%%%  

\subsection{Final states }

The final states which contain a RHN, $N_i$ are of our interest at the LHC. Due to singlet nature of RHN, it is difficult to produce them directly at the colliders viz., at the LHC. Thus such states can arise from either the decays of heavy neutral Higgs bosons $H$, the pseudoscalar $A$, or from the decays of the charged Higgs boson $H^\pm$.
The heavy neutral Higgs boson dominantly decays to $N_i \nu_i$ and the light pseudoscalar decays to $\tau\bar{\tau}$ and $N_i\nu_i$ depending on the available phase space.  

The associated production of heavy Higgs boson along with pseudoscalar can have interesting decay topology as given in Eq.~\ref{top1}. Given the mass spectrum for BP1 in Table~\ref{bps}, the heavy Higgs can decay to $N_i \nu_i$ and the light pseudoscalar dominantly decays to tau anti-tau pair giving rise to di-tau plus opposite sign dilepton (OSD) final states as shown in Eq.~\ref{top1}, where the leptons can be of different flavours. Thus it would be easy to distinguish the final state from the $Z$ boson contamination for the di-lepton. 
\bea\label{top1}
p p &\to & A H \to \tau \bar{\tau} N_i \nu_i  \nn \\
&\to&  \tau \bar{\tau} W^\pm \ell^\mp_i \nu_i \nn \\
&\to &  \tau \bar{\tau} \ell^\pm_j \nu_j \ell^\mp_i \nu_i ,
\eea
where $\ell^\pm_{i,j} = e^\pm , \,\mu^\pm,\, \tau^\pm$. 

However, our main focus in this article is to probe the charged Higgs boson via its decay mode comprised of RHN, $N$. The light charged Higgs boson  decays in the following kinematically allowed  final states, 
\bea\label{chdcy}
H^\pm &\to &\tau \nu \nn \\
&&  e N \nn \\
&& A W^\pm .
\eea

If $m_{H^\pm} > m_N$, then the produced charged Higgs can decay to $\ell^\pm N$. Such RHN further decays via two-body or three-body decay to leptons and gauge bosons or leptons and jets, respectively. Thus for Type-X, where a very light charged Higgs boson is still allowed from the current LHC bounds \cite{chLHC} unlike the Type-II charged Higgs boson, we can explore such light charged Higgs boson by searching the final states given below in Eqs.~\ref{top2},~\ref{fs1}, at the LHC. In this case, the dominant production mode is $p\,p \to A H^\pm$, where the charged Higgs boson further decays into $N_i \ell^\pm_i$ given as 
\bea\label{top2}
p p &\to & A H^\pm \to \tau \bar{\tau} N_i \ell^\pm_i  \nn \\
&\to&  \tau \bar{\tau} W^\pm \ell^\mp_i \ell^\pm_i \nn \\
&\to &  \tau \bar{\tau} \ell^\pm_j \nu_j \ell^\mp_i \ell^\pm_i
\eea
where $\ell^\pm_{i,j} = e^\pm ,\, \mu^\pm,\, \tau^\pm$. 
In collider only electron or muon can be detected as stable charged leptons giving rise to the following final state
\be\label{fs1}
p p \to   2 \tau + 2e(2\mu)  +\mu (e) + \ptmiss .
\ee

The charged Higgs if decays to electron and RHN then it can give rise to  signatures with different lepton flavours in final states as in the next step the
RHN further decays to $e^\mp W^\pm,\, Z \nu,\, h \nu$. As a result, we can have $2\tau_{\rm{jet}} +2W^\pm  +2e^\mp$ or $2\tau_{\rm{jet}}+W^\pm +e^\mp + (\ell^+ \ell^-)$. The interesting point to see that the gauge bosons decays to leptons via gauge coupling and so do not violate lepton flavours. Depending on the decays 
of RHN,  we can have multi-leptonic final states with lepton flavour violation.

For the searches of single charged Higgs boson, the $bg$ fusion is still dominant \cite{sch,NMSSMCH}. In our case however, the final state lepton(s) can have different flavours ($e,\, \mu $) owing to different branching ratios of Higgs boson to $e N$ and $\mu N$ due to non-democratic Yukawa for BP4 . 

\section{Collider simulation at the LHC}
\label{sec:simulation}

For the chosen benchmark points we will focus on these non-standard decays of the charged Higgs boson as well as the other Higgs bosons. We use  CalcHEP  to calculate the cross-sections and the decay branching fractions from the benchmark points. The `lhe' events are generated  and fed to {\tt PYTHIA} \cite{pythia}  for hadronization and fragmentation via the `lhe' interface \cite{lhe}. The simulation at hadronic level has been performed using the {\tt Fastjet-3.0.3} \cite{fastjet} with the {\tt CAMBRIDGE AACHEN} algorithm. We have selected a jet size $R=0.5$ for the jet formation, with the following criteria:

\begin{itemize}
	\item the calorimeter coverage is $\rm |\eta| < 4.5$
	
	\item the minimum transverse momentum of the jet $ p_{T,min}^{jet} = 10$ GeV and jets are ordered in $p_{T}$
	\item leptons ($\rm \ell=e,~\mu$) are selected with
	$p_T \ge 10$ GeV and $\rm |\eta| \le 2.5$
	\item no jet should be accompanied by a hard lepton in the event
	\item $\Delta R_{\ell j}\geq 0.4$ and $\Delta R_{\ell \ell}\geq 0.2$
	\item Since an efficient identification of the leptons is crucial for our study, we additionally require  
	a hadronic activity within a cone of $\Delta R = 0.3$ between two isolated leptons to be $\leq 0.15\, p^{\ell}_T$ GeV, with 
	$p^{\ell}_T$ being the transverse momentum of the lepton, in the specified cone.
	
\end{itemize}
%%%%%%%%%%%%%%%%%%%%%%%%%%%

%%%%%%%%%%%%%%%%%%%%%%%%%%%%%%%%%%%%%%%%%%%%%%%%%%%%%%%%
\begin{figure}[hbt]
	\begin{center}
		\includegraphics[width=0.33\linewidth, angle=-90]{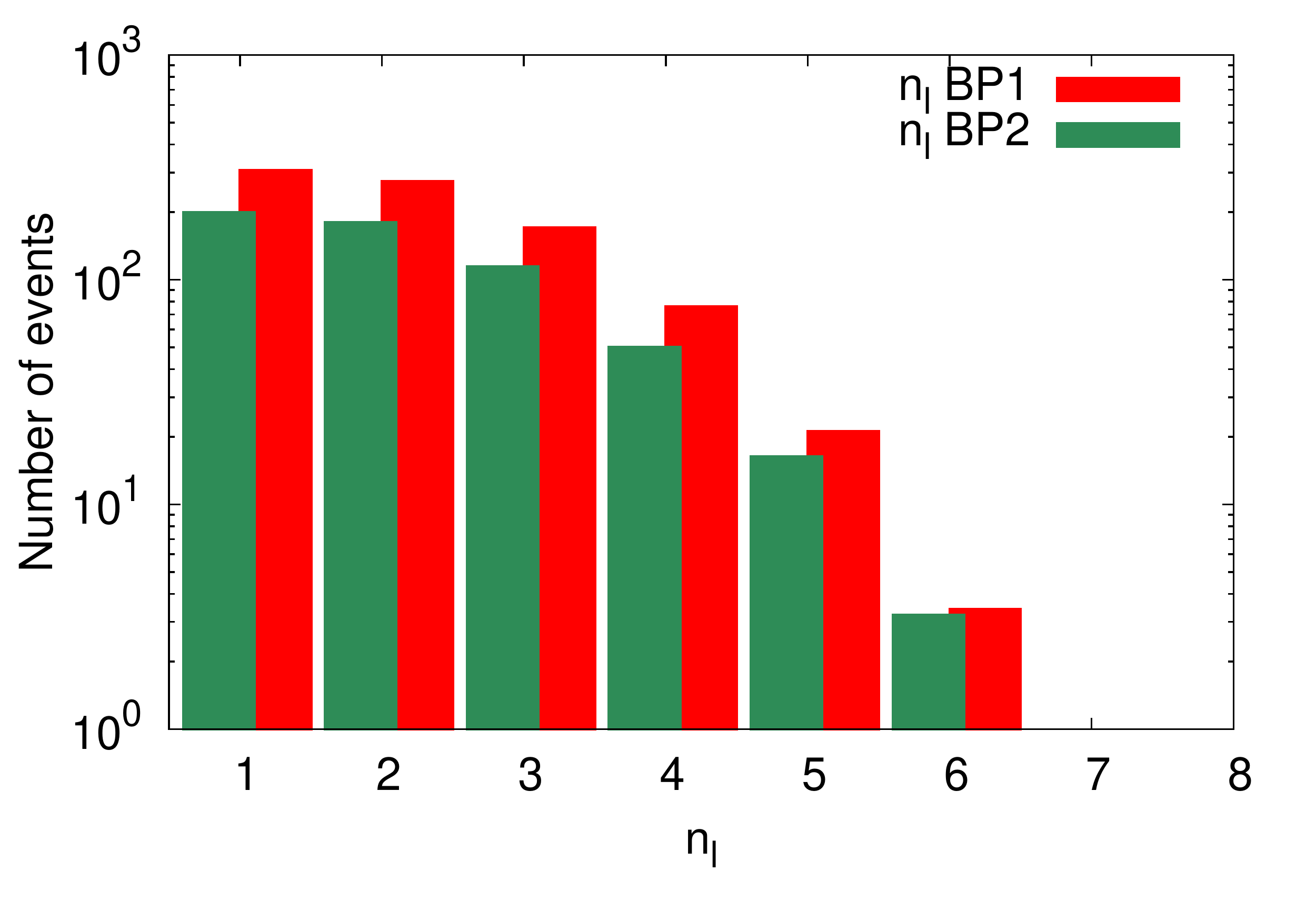}
		\includegraphics[width=0.33\linewidth, angle=-90]{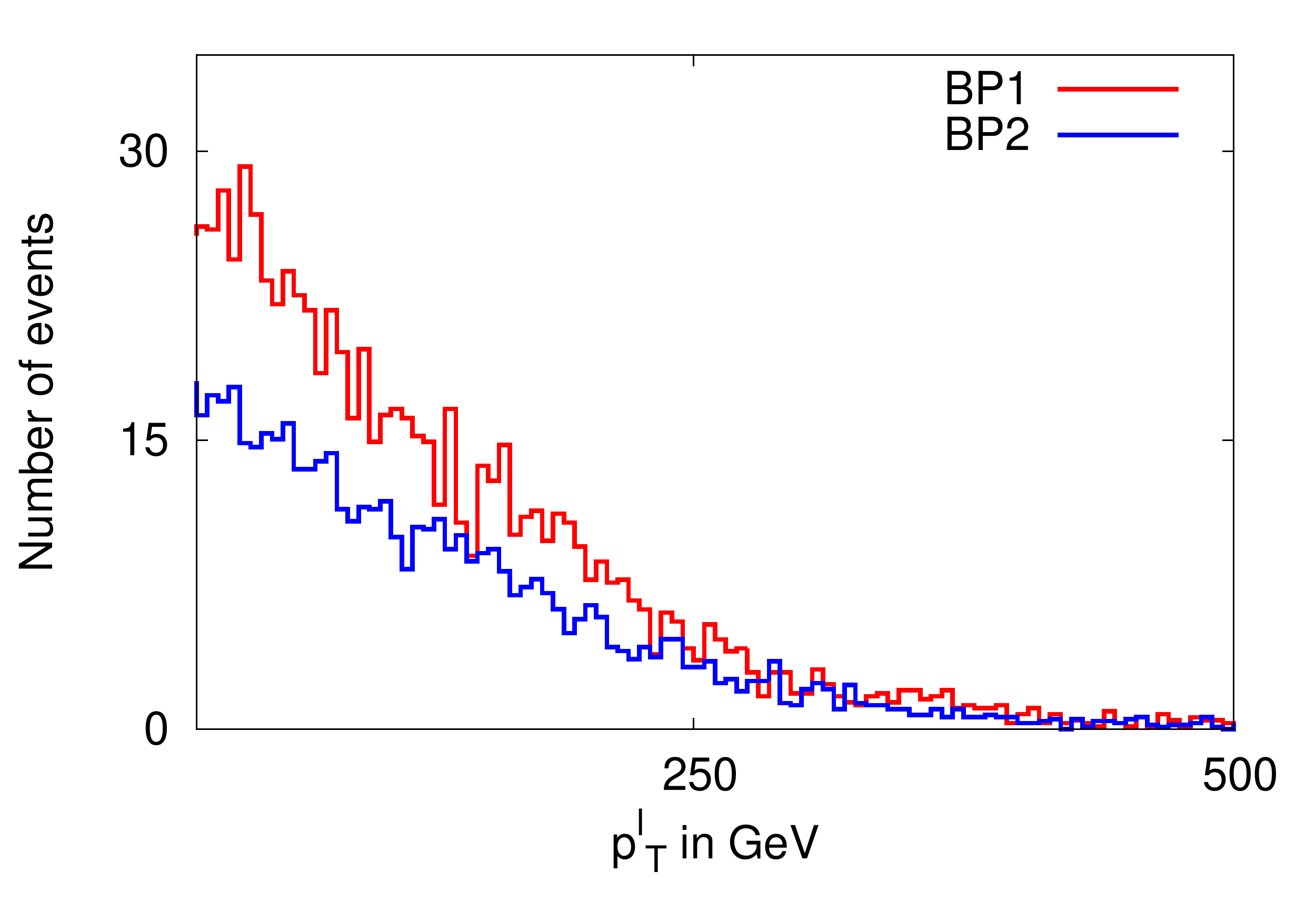}
		\caption{ $n_{\ell}$ distribution (left panel) and $p^{\ell}_T$ distribution (right panel) for BP1 and BP2 at an integrated luminosity of 100\,fb$^{-1}$ at the LHC with center of mass energy of 14\,TeV.}\label{dis1}
	\end{center}
\end{figure}
%%%%%%%%%%%%%%%%%%%%%%%%%%%%%%%%%%%%%%%%%%%%%%%%%%%%%%%
\begin{figure}[hbt]
	\begin{center}
		\includegraphics[width=0.33\linewidth, angle=-90]{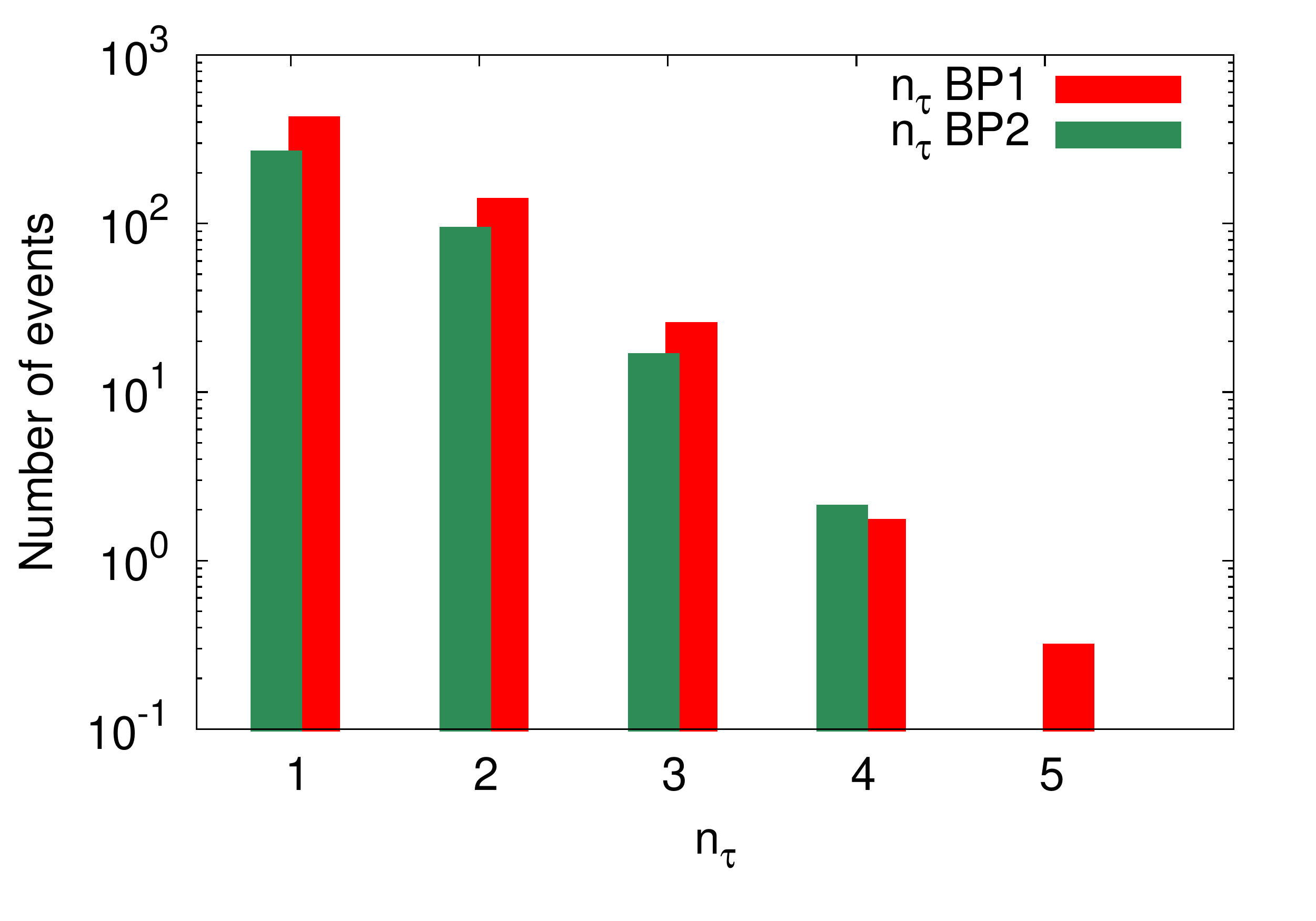}
		\includegraphics[width=0.33\linewidth, angle=-90]{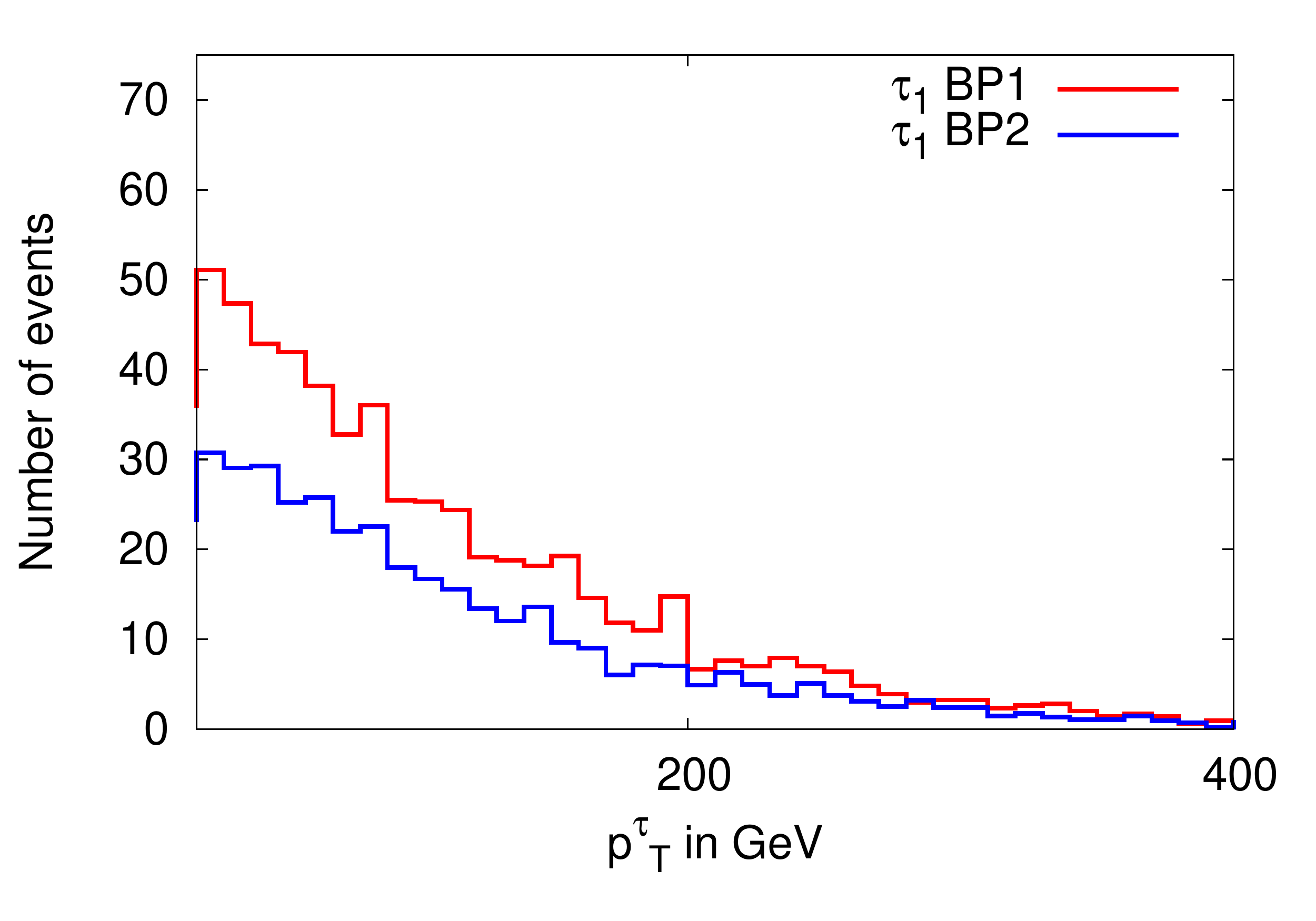}
		\caption{ $n_{\tau_{\rm{jet}}}$ distribution (left panel) and $p^{\tau}_T$ distribution (right panel) for BP1 and BP2 at an integrated luminosity of 100\,fb$^{-1}$ at the LHC with center of mass energy of 14\,TeV.}\label{dis2}		
	\end{center}
\end{figure}

%%%%%%

Equipped with the above set up and cuts we plot the lepton multiplicity $n_{\ell}$ and $p_T$ distribution in Fig.~\ref{dis1}. Here the production process for the benchmark points is $pp \to H^+ H^-$. Such $H^\pm$ can decay to $\ell^\pm N$ and the final state can have maximum of six charged leptons with non-universal lepton flavour number depending on the non-democratic Yukawa coupling $Y_{N_i}$. Fig.~\ref{dis1} (left panel)  depicts that we can tag those multi-leptons  as isolated charged leptons. In Fig.~\ref{dis1} (right panel) we show the $p^{\ell}_T$ distribution and some of them can actually be hard, as they may originate from the decay of the charged Higgs boson. Then there are relatively soft leptons arising from the $W^\pm$ decays. Finally the most soft charged leptons will come from the decay of the RHN $N_i$ due to smaller phase space for the decays to $\ell^\pm W^\mp,\, Z\nu$ states.

Figure~\ref{dis2} describes the tau multiplicity $n_{\tau_{\rm{jet}}}$ and $p^\tau_T$ distribution in left and right panels, respectively. The main source of the taus are from the decay of the pseudoscalar boson. The charged Higgs boson has sufficiently large branching fraction to $AW^\pm$ for BP1, BP3 and BP4, which can  give rise to multi-tau signature along with the taus coming from the decays of the gauge bosons. For the $p^\tau_T$ distribution in Fig.~\ref{dis2}, we only plot the events arising from the charged Higgs pair production. In the analysis we have considered all the production modes. The taus here are detected as hadronic tau jets $\tau_{\rm{jet}}$ \cite{taujet,cmstdr}. The taus coming from the pseudoscalar can be hard depending on the mass of the pseudoscalar which can be noticed from the right panel. 

\section{Results}
\label{sec:results}
In this section we present the event numbers for the final states for the benchmark points along with the dominant SM backgrounds.  We focus on multi-tau and multi-lepton final states in which we also tag the lepton flavours in order to probe the inverse seesaw Yukawa coupling $Y_{N_{i}}$. In the first few subsections we discuss the results for BP1, BP2 and BP4, and the phenomenology for BP3 is discussed separately in subsection \ref{sec:LightA} due to the presence of light pseudoscalar boson.

\subsection{$2\tau_{\rm{jet}} + 2\ell$}
Table~\ref{2tau2l} presents the number of events for $2\tau_{\rm jet} +2\ell$, $2\tau_{\rm jet} +2e$, $2\tau_{\rm jet} + 2\mu$ and $2\tau_{\rm jet} +1 e + 1 \mu$ respectively at the LHC with an integrated luminosity of 100\,fb$^{-1}$. For the SM backgrounds we have considered all possible potential backgrounds in the analysis and only the non-zero ones are listed in the table. To be explicit, we calculated the following cases;  $t\bar{t}$, $t\bar{t}Z$, $t\bar{t}W^\pm$, $tZW^\pm$, $VV$ and $VVV$, where $V\in\{ Z, W^\pm\}$ with all combinations.

%%%%%%%%%%%%%%%%% 2tau +2l %%%%%%%%%%%%%%%%%%
\begin{table}[thb]
	\begin{center}
		\hspace*{-1.cm}
		\renewcommand{\arraystretch}{1}
		\begin{tabular}{||c|c||c|c|c|c||c|c|c|c||}
			\hline\hline
			\multicolumn{2}{||c||}{\multirow{3}{*}{Final states}}&\multicolumn{4}{|c||}{Benchmark Points}&\multicolumn{4}{|c||}{Backgrounds}\\
			\cline{3-10}
			\multicolumn{2}{||c||}{}&\multirow{2}{*}{BP1}&\multirow{2}{*}{BP2}&\multirow{2}{*}{BP3}&\multirow{2}{*}{BP4}&\multirow{2}{*}{$t\bar t$}&\multirow{2}{*}{$t\bar t V$}&\multirow{2}{*}{$tZW^\pm$}&$VV/$ \\
			\multicolumn{2}{||c||}{}&&&&&&&&$VVV$ \\
			\hline\hline
				\multirow{4}{*}{\rotatebox[origin=c]{90}{$2\tau_{\rm{jet}} + 2\ell$}}&$H A$&240.7&64.9&207.6&258.8&\multirow{4}{*}{164.8}&\multirow{4}{*}{11.3}&\multirow{4}{*}{3.5}&\multirow{4}{*}{632.7}\\
		&$H^\pm H^\mp$&22.8&23.6&13.7&53.0&&&&\\
			&$H H^\pm$&182.2&53.6&267.5&131.6&&&&\\
			&$A H^\pm$&569.7&180.0&168.7&460.5&&&&\\
			\hline
		\multicolumn{2}{|c|}{\multirow{2}{*}{Total} }&1015.4&322.2  &657.5  &903.9 &\multicolumn{4}{|c|}{\multirow{2}{*}{$812.2 \pm 121.8$}}\\
		\multicolumn{2}{|c|}{} & $\pm 152.3 $ &$\pm 48.3 $  &$\pm 98.7 $  &$\pm 135.6 $ &\multicolumn{4}{|c|}{ }\\
			\hline
				\multicolumn{2}{||c||}{$N_{sig}$(in $\sigma$)}&$\{20.4,27.1\}$&$\{7.9,11.4\}$ &$\{14.5,19.9\}$&$\{18.6,25.0\}$&\multicolumn{4}{|c||}{}\\
				\hline\hline
				\multirow{4}{*}{\rotatebox[origin=c]{90}{$2\tau_{\rm{jet}} + 2e$}}&$H A$&69.1&16.9&54.6&76.8&\multirow{4}{*}{40.0}&\multirow{4}{*}{2.7}&\multirow{4}{*}{0.9}&\multirow{4}{*}{274.0}\\
			&$H^\pm H^\mp$&8.6&9.5&3.0&31.1&\multirow{3}{*}{}&\multirow{3}{*}{}&\multirow{3}{*}{}&\\
		&$H H^\pm$&63.1&21.1&78.9&74.0&&&&\\
		&$A H^\pm$&185.7&71.5&52.8&272.4&&&&\\
				\hline
				\multicolumn{2}{|c|}{\multirow{2}{*}{Total} } &326.5&119.0&189.3&454.3&\multicolumn{4}{|c|}{\multirow{2}{*}{$317.7 \pm 47.7$}}\\
				\multicolumn{2}{|c|}{} & $\pm 49.0 $ &$\pm 17.9 $  &$\pm 28.4 $  &$\pm 68.2 $ &\multicolumn{4}{|c|}{ }\\
				\hline
			\multicolumn{2}{||c||}{$N_{sig}$(in $\sigma$)}&$\{10.9,14.8\}$&$\{4.7,6.8\}$&$\{7.0,9.9\}$&$\{14.1,18.6\}$&\multicolumn{4}{|c||}{}\\
			\hline\hline
			\multirow{4}{*}{\rotatebox[origin=c]{90}{$2\tau_{\rm{jet}} + 2\mu$}}&$H A$&75.5&17.6&51.7&67.9&\multirow{4}{*}{42.0}&\multirow{4}{*}{4.7}&\multirow{4}{*}{1.4}&\multirow{4}{*}{328.7}\\
			&$H^\pm H^\mp$&9.0&9.7&3.7&9.3&\multirow{3}{*}{}&\multirow{3}{*}{}&\multirow{3}{*}{}&\\
			&$H H^\pm$&69.0&22.7&78.6&26.8&&&&\\
			&$A H^\pm$&195.6&78.8&46.3&76.2&&&&\\
				\hline
				\multicolumn{2}{|c|}{\multirow{2}{*}{Total} } &349.1&128.8&180.3&180.2&\multicolumn{4}{|c|}{\multirow{2}{*}{$376.8 \pm 56.5$}}\\
				\multicolumn{2}{|c|}{} & $\pm 52.4 $ &$\pm 19.3 $  &$\pm 27.0 $  &$\pm 27.0 $ &\multicolumn{4}{|c|}{ }\\
				\hline
			\multicolumn{2}{||c||}{$N_{sig}$(in $\sigma$)}&$\{11.0,14.9\}$&$\{4.7,6.8\}$&$\{6.3,9.0\}$&$\{6.3,9.0\}$&\multicolumn{4}{|c||}{}\\
			\hline\hline
			\multirow{4}{*}{\rotatebox[origin=c]{90}{$~~2\tau_{\rm{jet}}\! +\! e \!+\!\mu~$}}&$H A$&110.9&30.4&103.6&128.5&\multirow{4}{*}{82.4}&\multirow{4}{*}{3.8}&\multirow{4}{*}{1.3}&\multirow{4}{*}{30.4}\\
				&$H^\pm H^\mp$&8.1&8.3&7.5&20.7&&&&\\
			&$H H^\pm$&81.0&16.8&148.8&51.2&&&&\\
			&$A H^\pm$&251.3&56.6&85.8&167.6&&&&\\ 	%\noalign{\vskip1pt}
			\hline\hline 		
			\multicolumn{2}{|c|}{\multirow{2}{*}{Total}} &451.4&112.1&345.7&368.0&\multicolumn{4}{|c|}{\multirow{2}{*}{$118.0 \pm 17.7$}}\\
			\multicolumn{2}{|c|}{} & $\pm 67.7 $ &$\pm 16.8 $  &$\pm 51.9 $  &$\pm 55.2 $ &\multicolumn{4}{|c|}{ }\\
			\hline
			\multicolumn{2}{||c||}{$N_{sig}$(in $\sigma$)}&$\{16.8,20.9\}$&$\{6.3,8.5\}$&$\{14.2,17.8\}$&$\{14.8,18.5\}$&\multicolumn{4}{|c||}{}\\
			%\cline{1-5}
		%	\multicolumn{2}{||c||}{$\mathcal{L}_{5\sigma}$ (fb$^{-1}$)}&$\gg3000$&$\gg3000$&$\gg3000$&437&\multicolumn{4}{|c||}{}\\
			\hline\hline
		\end{tabular}
		\caption{The number of events for $2\tau_{\rm{jet}}+2\ell $ final state at 100\,fb$^{-1}$ of integrated luminosity at the LHC with 14\,TeV center of mass energy. The range for $N_{sig}$ is calculated incorporating the systematic uncertainties in signal and background events as well.  The flavour tagging ($e, \mu$) has been implemented.}\label{2tau2l}
	\end{center}
\end{table}
%%%%%%%%%%%%%%%%%%%%%%%%%%%%%%%%%%%%%%%%%%%%%
The finalstate is reached in $\tau$ and we tag such $\tau$s hadronically as $\tau_{\rm{jet}}$ \cite{taujet,cmstdr}. Here, in the case of the $\tau_{\rm{jet}}$ we have considered the hadronic decay of the $\tau$ to be characterized by at least one charged track with $\Delta R \leq 0.1$ of the
candidate $\tau_{\rm{jet}}$ \cite{taujet,cmstdr}. The demand of such hadronically reconstructed $\tau_{\rm{jet}}$ along with the criteria of two isolated leptons reduce the SM background drastically. Given the finalstates with multi-leptons, $t\bar{t}$ and $t\bar{t}W^\pm$  seem to fail to contribute as backgrounds and  the major contributions are expected to come from the di- and triple-gauge boson production including the $Z$ boson.  However, mis-tagging of normal jets as tau-jets can contribute as SM backgrounds; especially for $t\bar{t}$ due it's large cross-section. For the completeness of the analysis we have considered a mis-tagging efficiency of $2\%$, which is a conservative estimate for large $p_T$ tau-jets \cite{mistau}. The finalstates $2\tau+2\ell$ (in Table~\ref{2tau2l}) and  $2\tau +3\ell$ (in Table~\ref{2tau3l})  are affected by the mis-tagging efficiency. However, in Table~\ref{4tau} such changes are insignificant.

The signal and the background numbers are subject to the uncertainties arising from the systematics as well as the statistics.  Here we mainly focus on the systematics uncertainties and predict the range for signal significance in the succeeding paragraphs. The uncertainty in the cross-section is dominated by the PDF uncertainty which is around 10\%, then the jet-scale uncertainty is considered as 3\%  \cite{cmstdr} and the tau-jet mis-tagging uncertainty is taken to be 8.8\% \cite{mistau}. In Table~\ref{2tau2l} and Table~\ref{2tau3l} the event numbers are given with their uncertainties for both the signal and backgrounds.

As mentioned earlier, for the considered benchmark points, dominant contribution arises from $AH^\pm$ production but other production processes are also significant.  We see that for $2\tau_{\rm{jet}} + 2\ell$ channel, the minimum reach of BP1, BP2 and BP4 are $20.4\sigma,\, 7.9\sigma$ and $18.6\sigma$, respectively.  The signal significance denoted by $N_{sig}$ is calculated in a conservative approach as signal/$\sqrt{\rm signal + background}$.  

The demand of only electron flavour can probe the non-democratic inverse seesaw Yukawa coupling $Y_{N_i}$ scenario. The final state of $2\tau_{\rm{jet}} + 2e$ reduces both the signal as well as the background numbers. The signal significance for the benchmark points reduces  to $10.9\sigma,\, 4.7\sigma$ and $14.1\sigma$ respectively for BP1, BP2 and BP4.  

Next we look at the final state having $2\tau_{\rm{jet}} + 2\mu$ where for BP1 and BP2 have event number similar to $2\tau_{\rm{jet}} + 2e$ channel as they have democratic inverse seesaw Yukawa coupling $Y_{N_i}$. However, in BP4, the number of event for $2\tau_{\rm{jet}} + 2\mu$ reduces substantially due to non-democratic choice $Y_{N_1}=0.5, \, Y_{N_{2,3}}=0.1$.  The charged Higgs boson as well  CP-even heavy Higgs boson decay to $AW^\pm$ and $AZ$ for BP4, which contributes to di-muon final state.  The respective minimum signal significance for BP1, BP2 and BP4 are $11.0\sigma, 4.7\sigma$ and $6.3\sigma$,  which is lower only for BP4 with respect to the $2\tau_{\rm{jet}} + 2e$ final state. 

Finally we also present the event numbers for $2\tau_{\rm{jet}} + 1e + 1\mu$ final states and the corresponding minimum signal significances are $16.8\sigma, 6.3\sigma$ and $14.8\sigma$ for BP1, BP2 and BP4, respectively. 

%%%%%%%%%%%%%% Signal Significance %%%%%%%%%%%%%%%%%%
\begin{figure}[t]
	\begin{center}
		\includegraphics[width=0.34\linewidth, angle=-90]{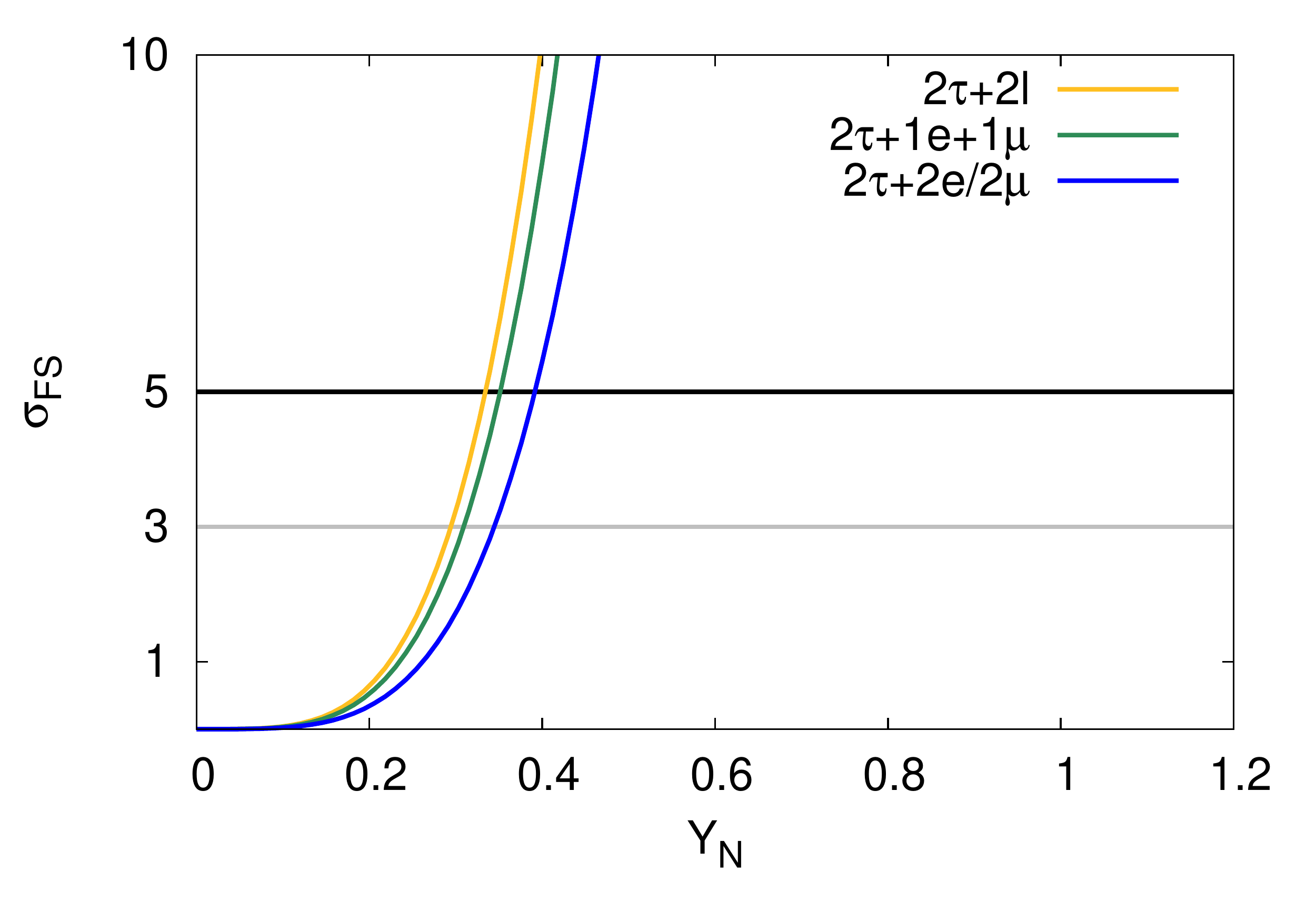} 	\hspace*{0.2cm}
		\includegraphics[width=0.34\linewidth, angle=-90]{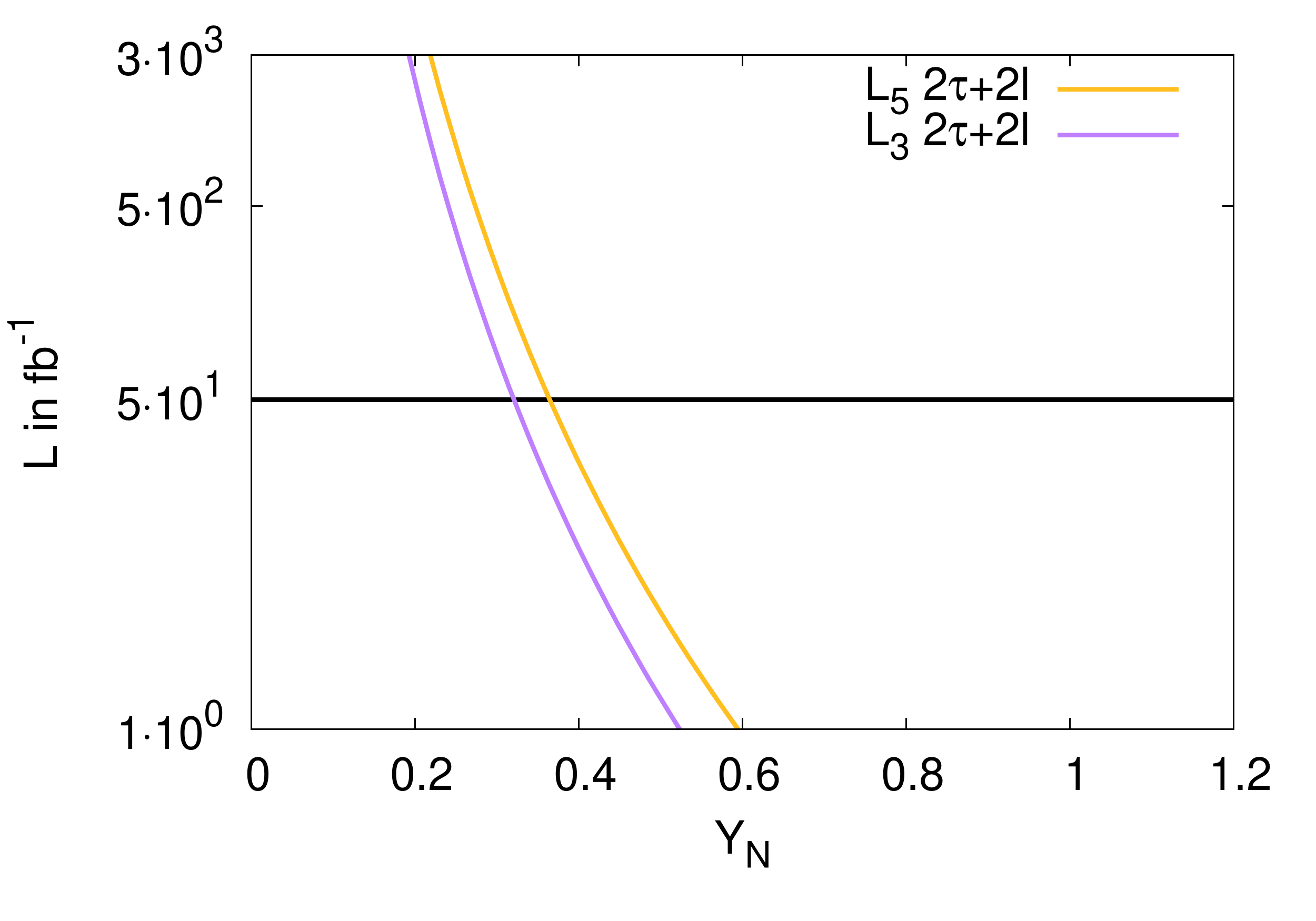}
		\caption{(Left panel) We present the signal significance verses $Y_N$ for the chosen final states at an integrated luminosity of 100\,fb$^{-1}$. (Right panel) The required luminosity for $5\sigma$ signal significance ($L_5$) verses $Y_N$ are shown for $2\tau+2\ell$ final state  with center of mass energy of 14\,TeV at the LHC.}\label{lum}
	\end{center}
\end{figure}

It can be seen from the above discussion that $2\tau+ 2\ell$ final states have very high signal significance for all BPs. We use these modes to explore the reach to probe Yukawa coupling $Y_N$ at the LHC with center of mass energy of 14\,TeV. The result is depicted in Fig.~\ref{lum}. The left panel shows the variation of signal significances w.r.t. the Yukawa coupling $Y_N$,  where purple, green and blue bands correspond to the $2\tau+ 2\ell$, $2\tau+ 1e+1\mu$ and $2\tau+ 2e/2\mu$ final states respectively for an integrated luminosity of 100\,fb$^{-1}$. The horizontal gray line corresponds to the signal significance of $3\sigma$ over SM backgrounds, whereas the black line corresponds to $5\sigma$ significance. It is evident that the inclusive $2\tau+2\ell$ has the maximum signal significance and the RHN Yukawa coupling $Y_N \gsim 0.3$  (within 15\% systematic uncertainty as shown in the bands) for inverse seesaw can be probed with early data.

The Table~\ref{2tau2l} result is then used to obtain the contour plots in Fig.~\ref{lum} right panel  for the signal significance in the plane spanned by integrated luminosity and the inverse seesaw Yukawa coupling $Y_N$.  Here we present the contours of $3\sigma$ and $5\sigma$ significance by green and red bands, respectively, for the signal $2\tau+2\tau$ at the LHC with 14\,TeV center of mass energy, in the integrated luminosity verse Yukawa coupling $Y_N$ plane. We can see that,  within 15\% systematic uncertainty, $Y_N \sim 0.5$ and $\sim 0.6$ can receive $3\sigma$ and $5\sigma$ discovery respectively. For lower values of $Y_N$ we need higher integrated luminosity. For $\mathcal{O}(100)$\,GeV RHN mass, the LHC at 3000\,fb$^{-1}$ can probe the inverse seesaw Yukawa coupling $\sim 0.2$.

\subsection{$2\tau_{\rm{jet}} + 3\ell$}

Motivated by the topologies as described in Eq.~\ref{top1} and in Eq.~\ref{top2} we look for $3\ell$ final state in association with $2\tau_{\rm{jet}} $. Obviously, the demand of $3\ell$ reduces the SM backgrounds to negligible order. Table~\ref{2tau3l} present the number of events at the LHC with 14\,TeV center of mass energy at an integrated luminosity of 100\,fb$^{-1}$. 

The inclusive $2\tau_{\rm{jet}} +3\ell$ final state has a minimum signal significance of $12\sigma, 7.5 \sigma$ and $11.6\sigma$ respectively for BP1, BP2 and BP4. $2\tau_{\rm{jet}} +2e +1\mu$
signal has significance of $6.9\sigma, 3.8\sigma$ and $7.6\sigma$ respectively for BP1, BP2 and BP4.  If we tag events with di-muon we find $2\tau_{\rm{jet}} +2\mu +1e$ with signal significance of $7.0\sigma, 4.1\sigma$ and $5.1\sigma$ respectively for BP1, BP2 and BP4. As before for BP4, the significance drops down from the $2e$ case due to non-democratic inverse seesaw Yukawa $Y_{N_i}$.  Such scenario can lead to experimental signature of lepton flavour violation in the final states \cite{pbiss, uNs, lfv}. 
%%%%%%%%%%%%%%%%% 2tau +3l %%%%%%%%%%%%%%%%%%
\begin{table}[thb]
	\begin{center}
		\hspace*{-0.5cm}
		\renewcommand{\arraystretch}{1}
		\begin{tabular}{||c|c||c|c|c|c||c|c|c|c||}
			\hline\hline
			\multicolumn{2}{||c||}{\multirow{3}{*}{Final states}}&\multicolumn{4}{|c||}{Benchmark Points}&\multicolumn{4}{|c||}{Backgrounds}\\
			\cline{3-10}
			\multicolumn{2}{||c||}{}&\multirow{2}{*}{BP1}&\multirow{2}{*}{BP2}&\multirow{2}{*}{BP3}&\multirow{2}{*}{BP4}&\multirow{2}{*}{$t\bar t$}&\multirow{2}{*}{$t\bar t V$}&\multirow{2}{*}{$tZW^\pm$}&$VV/$ \\
			\multicolumn{2}{||c||}{}&&&&&&&&$VVV$ \\
			\hline\hline
			\multirow{4}{*}{\rotatebox[origin=c]{90}{$2\tau_{\rm{jet}} + 3\ell$}}&$H A$&24.2&0.0&4.8&23.9&\multirow{4}{*}{0.2}&\multirow{4}{*}{0.1}&\multirow{4}{*}{0.0}&\multirow{4}{*}{0.9}\\
			&$H^\pm H^\mp$&5.6&6.4&0.9&13.5&&&&\\
			&$H H^\pm$&43.0&13.5&50.2&32.8&&&&\\
			&$A H^\pm$&97.8&47.0&21.1&91.0&&&&\\
			\hline
			\multicolumn{2}{|c|}{\multirow{2}{*}{Total} }&170.6&66.8&77.0&161.1 &\multicolumn{4}{|c|}{\multirow{2}{*}{$1.2\pm 0.2$}}\\
			\multicolumn{2}{|c|}{} & $\pm 25.6 $ &$\pm 10.0$  &$\pm 11.6 $  &$\pm 24.2 $ &\multicolumn{4}{|c|}{ }\\
			\hline
			\multicolumn{2}{||c||}{$N_{sig}$(in $\sigma$)}&$\{12.0,14.0\}$&$\{7.5,8.7\}$ &$\{8.0,9.4\}$&$\{11.6,13.6\}$&\multicolumn{4}{|c||}{}\\
			\hline\hline
			\multirow{4}{*}{\rotatebox[origin=c]{90}{$2\tau_{\rm{jet}} \!+\! 2e\!+\!\mu$}}&$H A$&6.6&0.00&1.7&6.9&\multirow{4}{*}{0.1}&\multirow{4}{*}{0.0}&\multirow{4}{*}{0.0}&\multirow{4}{*}{0.5}\\
			&$H^\pm H^\mp$&1.3&2.4&0.3&6.2&&&&\\
			&$H H^\pm$&13.9&2.8&19.0&13.1&&&&\\
			&$A H^\pm$&35.1&12.6&9.9&41.7&&&&\\
			\hline
			\multicolumn{2}{|c|}{\multirow{2}{*}{Total} } &53.3&17.8&30.9&67.9&\multicolumn{4}{|c|}{\multirow{2}{*}{$0.6 \pm 0.1$}}\\
			\multicolumn{2}{|c|}{} & $\pm 8.5 $ &$\pm 2.7 $  &$\pm 4.6 $  &$\pm 10.2 $ &\multicolumn{4}{|c|}{ }\\
			\hline
			\multicolumn{2}{||c||}{$N_{sig}$(in $\sigma$)}&$\{6.9,8.1\}$&$\{3.8,4.4\}$&$\{5.1,5.9\}$&$\{7.6,8.8\}$&\multicolumn{4}{|c||}{}\\
			\hline\hline
			\multirow{4}{*}{\rotatebox[origin=c]{90}{$2\tau_{\rm{jet}} \!+\! 2\mu\!+\!e~$}}&$H A$&8.1&0.0&0.8&7.4&\multirow{4}{*}{0.1}&\multirow{4}{*}{0.0}&\multirow{4}{*}{0.0}&\multirow{4}{*}{0.2}\\
			&$H^\pm H^\mp$&1.6&1.6&0.3&1.9&&&&\\
			&$H H^\pm$&17.1&4.1&19.8&7.2&&&&\\
			&$A H^\pm$&31.5&14.3&6.3&14.2&&&&\\  \noalign{\vskip2pt}
			\hline
			\multicolumn{2}{|c|}{\multirow{2}{*}{Total} } &58.3&20.0&27.2&30.8&\multicolumn{4}{|c|}{\multirow{2}{*}{$0.3 \pm 0.0$}}\\
			\multicolumn{2}{|c|}{} & $\pm 8.7 $ &$\pm 3.0 $  &$\pm 4.1 $  &$\pm 4.6 $ &\multicolumn{4}{|c|}{ }\\
			\hline
			\multicolumn{2}{||c||}{$N_{sig}$(in $\sigma$)}&$\{7.0,8.2\}$&$\{4.1,4.8\}$&$\{4.8,5.6\}$&$\{5.1,5.9\}$&\multicolumn{4}{|c||}{} \\
			\hline\hline
		\end{tabular}
		\caption{The number of events for $2\tau_{\rm{jet}} +3\ell $ final state at 100\,fb$^{-1}$ of integrated luminosity at the LHC with 14\,TeV center of mass energy. The range for $N_{sig}$ is calculated incorporating the systematic uncertainties in signal and background events as well. The flavour tagging ($e, \mu$) has been implemented.}\label{2tau3l}
	\end{center}
\end{table}

%%%%%%%%%%%%%%%%%%%%%%%%%%%%%%%%%%%%%%%%%%
\subsection{Very light  pseudoscalar}
\label{sec:LightA}

As a consequent of very light pseudoscalar Higgs  boson ($m_A \sim 50$ GeV), BP3 possess very different phenomenology compared to the other three benchmark points as the RHN completely decays to light pseudoscalar and light neutrinos (Table~\ref{brN}).  The $H^\pm$ and $H$ contribute to the RHN final states  with  $\sim 40\%$ branching ratio. 
The final states searched in the previous subsections namely $2\tau_{\rm{jet}} +2\ell$ and $2\tau_{\rm{jet}} +3\ell$ also provide quite reasonable significance for BP3 as can be noted from Table~\ref{2tau2l} and Table~\ref{2tau3l}, respectively, for all channels. Apart from these modes, we can also explore the final states comprised of RHN, with the topologies given in Eqs.~\eqref{bp3fs}--\eqref{bp3fs2}.

%
%%%%%%%%%%%%%%%%% 4tau + X %%%%%%%%%%%%%%%%%%
\begin{table}[thb]
	\begin{center}
		%	\hspace*{-1.75cm}
		\renewcommand{\arraystretch}{1}
		\begin{tabular}{||c|c||c|c|c|c||c|c||}
			\hline\hline
			\multicolumn{2}{||c||}{\multirow{2}{*}{Final states}}&\multicolumn{4}{|c||}{Benchmark Points}&\multicolumn{2}{|c||}{Backgrounds}\\
			\cline{3-8}
			\multicolumn{2}{||c||}{}&BP1&BP2&BP3&BP4&$t\bar t\, V$&$VV/VVV$\\
			\hline\hline
			\multirow{3}{*}{$4\tau_{\rm{jet}} + \ptmiss$}&$H A$&104.6&21.4&78.5&94.4&\multirow{4}{*}{0.2}&\multirow{4}{*}{17.1}\\
			\multirow{3}{*}{$ \geq 30 \rm{GeV}$}&$H^\pm H^\mp$&1.0&1.8&2.2&6.3&\multirow{3}{*}{}&\multirow{3}{*}{}\\
			&$H H^\pm$&53.0&5.2&71.4&22.9&&\\
			&$A H^\pm$&142.8&14.6&45.1&80.7&&\\
			\hline
			\multicolumn{2}{|c|}{\multirow{2}{*}{Total}} &301.4&43.1&197.2&204.3&\multicolumn{2}{|c|}{\multirow{2}{*}{$17.3\pm 2.6$} }\\
			\multicolumn{2}{|c|}{} & $\pm 25.6 $ &$\pm 10.0$  &$\pm 11.6 $  &$\pm 24.2 $ &\multicolumn{2}{|c|}{}\\
			\hline
			\multicolumn{2}{||c||}{$N_{sig}$(in $\sigma$)}&$\{15.4,18.2\}$&$\{4.9,6.2\}$ &$\{12.2,14.6\}$&$\{12.5,14.9\}$&\multicolumn{2}{|c||}{}\\
			\hline\hline
			\multirow{4}{*}{$4\tau_{\rm{jet}} + 1e$}&$H A$&6.7&0.0&1.2&6.8&\multirow{4}{*}{0.0}&\multirow{4}{*}{0.0}\\
			&$H^\pm H^\mp$&0.4&0.1&0.3&1.1&\multirow{3}{*}{}&\multirow{3}{*}{}\\
			&$H H^\pm$&9.7&0.5&20.7&3.3&&\\
			&$A H^\pm$&22.7&2.6&7.3&14.0&&\\
			\hline
			\multicolumn{2}{|c|}{\multirow{2}{*}{Total}} &39.5&3.2&29.5&25.3&\multicolumn{2}{|c|}{\multirow{2}{*}{0.0}}\\
			\multicolumn{2}{|c|}{} & $\pm 5.9 $ &$\pm 0.5$  &$\pm 4.4 $  &$\pm 3.8 $ &\multicolumn{2}{|c|}{}\\
			\hline
			\multicolumn{2}{||c||}{limit(in $\sigma$)}&$1.9$&$1.2$ &$1.8$&$1.7$&\multicolumn{2}{|c||}{}\\
			\hline\hline
			\multirow{4}{*}{$4\tau_{\rm{jet}}+ 1\mu $}&$H A$&7.5&0.0&0.8&5.6&\multirow{4}{*}{0.0}&\multirow{4}{*}{0.0}\\
			&$H^\pm H^\mp$&0.1&0.4&0.4&2.0&\multirow{4}{*}{}&\multirow{4}{*}{}\\
			&$H H^\pm$&10.3&0.6&19.0&5.4&&\\
			&$A H^\pm$&22.5&2.1&6.1&13.2&&\\
			\hline
			\multicolumn{2}{|c|}{\multirow{2}{*}{Total}} &40.4&3.1&26.3&26.1&\multicolumn{2}{|c|}{\multirow{2}{*}{0.0}}\\
			\multicolumn{2}{|c|}{} & $\pm 6.1 $ &$\pm 0.5$  &$\pm 3.9 $  &$\pm 3.9 $ &\multicolumn{2}{|c|}{}\\
			\hline
			\multicolumn{2}{||c||}{limit(in $\sigma$)}&$1.9$&$1.2$ &$1.8$&$1.8$&\multicolumn{2}{|c||}{}\\
			\hline\hline
			\multirow{4}{*}{$4\tau_{\rm{jet}} + 2e$}&$H A$&1.3&0.0&0.2&1.6&\multirow{4}{*}{0.0}&\multirow{4}{*}{0.0}\\
			&$H^\pm H^\mp$&0.0&0.0&0.0&0.1&\multirow{3}{*}{}&\multirow{3}{*}{}\\
			&$H H^\pm$&0.5&0.0&0.1&0.3&&\\
			&$A H^\pm$&0.0&0.0&0.0&0.0&&\\
			\hline
			\multicolumn{2}{|c|}{Total} &1.8&0.0&0.3&2.0&\multicolumn{2}{|c|}{0.0}\\
			\hline
%			\multicolumn{2}{||c||}{limit(in $\sigma$)}&$1.9$&$1.2$ &$1.8$&$1.7$&\multicolumn{2}{|c||}{}\\
			\hline%\hline
			\multirow{4}{*}{$4\tau_{\rm{jet}}+ 2\mu $}&$H A$&2.0&0.0&0.2&1.4&\multirow{4}{*}{0.0}&\multirow{4}{*}{0.0}\\
			&$H^\pm H^\mp$&0.0&0.1&0.0&0.1&\multirow{3}{*}{}&\multirow{3}{*}{}\\
			&$H H^\pm$&1.2&0.0&0.3&0.3&&\\
			&$A H^\pm$&0.0&0.0&0.0&0.0&&\\
			\hline
			\multicolumn{2}{|c|}{Total} &3.2&0.1&0.5&1.8&\multicolumn{2}{|c|}{0.0}\\
			\hline
%			\multicolumn{2}{||c||}{limit(in $\sigma$)}&$1.9$&$1.2$ &$1.8$&$1.7$&\multicolumn{2}{|c||}{}\\
			\hline%\hline
			%				\multicolumn{2}{||c||}{$\mathcal{L}_{5\sigma}$ (fb$^{-1}$)}&$\gg3000$&$\gg3000$&$\gg3000$&437&\multicolumn{4}{|c||}{}\\
		\end{tabular}
		\caption{The number of events for $4\tau_{\rm{jet}} +\ptmiss \geq 30 \rm{GeV}$, $4\tau_{\rm{jet}} +1\ell$ and $4\tau_{\rm{jet}} +2\ell$  final states at 100\,fb$^{-1}$ of integrated luminosity at the LHC with 14\,TeV center of mass energy. The range for $N_{sig}$ is calculated incorporating the systematic uncertainties in signal and background events as well. The flavour tagging ($e, \mu$) has been implemented. }\label{4tau}
	\end{center}
\end{table}
%%%%%%%%%%%%%%%%%%%%%%%%%%%%%%%%%%%%%%%%%%
%\vspace*{-1.5cm}
\bea\label{bp3fs}
H^\pm H &\to &  N e^\pm N \nu \nn\\
& \to  & 2 A + e^\pm + 3\nu\nn \\
& \to  & 4\tau + e^\pm + \ptmiss \\
H^\pm H^\mp &\to&  N e^+ N e^- \nn\\
&\to & 4\tau + OSE  + \ptmiss \\
%\eea
%\bea
A H &\to & \tau \tau N\nu \nn\\
&\to & 4 \tau + \ptmiss\\
A H^\pm &\to& \tau \tau N e^\pm \nn \\
&\to & 4\tau + e^\pm + \ptmiss \label{bp3fs2}
\eea

The signal and non-zero background numbers are shown in Table~\ref{4tau} for all channels. For BP3 the RHNs decay completely to the $A\nu$ states, and further decay of $A$ to tau pairs enrich the $4\tau$ signature here. BP1, BP4 also compete with BP3 in these cases when produced in association with one pseudoscalar boson, which decays almost completely to tau pairs as well. We find $15.4\sigma,4.9\sigma,12.2\sigma$ and $12.5\sigma$ significance in  $4\tau_{\rm{jet}} + \ptmiss \geq 30 \rm{GeV}$ mode for BP1, BP2, BP3 and BP4, respectively. For $4\tau_{\rm{jet}} + 1e$ and $4\tau_{\rm{jet}} + 1\mu$ modes as no background events are observed, we use Poisson distribution to impose exclusion limits in the respective channels. It can be seen that  for BP1, BP3 and BP4 the  limits are just below $2\sigma$ level, however for BP2 these two contributions are suppressed as the pseudoscalar mostly decays to $N\nu$ states (with branching fraction 62\% given in Table~\ref{bra1}). 
%
%The signal significance is also very reasonable ($\ge5\sigma$) for $4\tau_{\rm{jet}} + 1e$ and $4\tau_{\rm{jet}} + 1\mu$ modes for BP1, BP3 and BP4, however contribution for BP2 is suppressed as the pseudoscalar mostly decays to $N\nu$ states (with branching fraction 62\% given in Table~\ref{bra1}). 
The channels with $4\tau_{\rm{jet}} + 2e$ and $4\tau_{\rm{jet}} + 2\mu$ are not at satisfactory level for 100\,fb$^{-1}$ luminosity  and we do not calculate the signal significance for these low signal event numbers and one needs to wait for more data for such prediction .

%%%%%%%%%%%%%%%%%%%%%%%%%%%%%%%%%%%%%%%%%%

\section{Reconstruction of charged Higgs boson mass}
\label{sec:Chiggs}

%%%%%%%%%%%%%%%%%%%%%%%%%%%%%%%%%%%%%%%%%%%%%%%%%%%%%%%%%%%%%%

\begin{figure}[hbt]
	\begin{center}
		\includegraphics[width=0.32\linewidth, angle=-90]{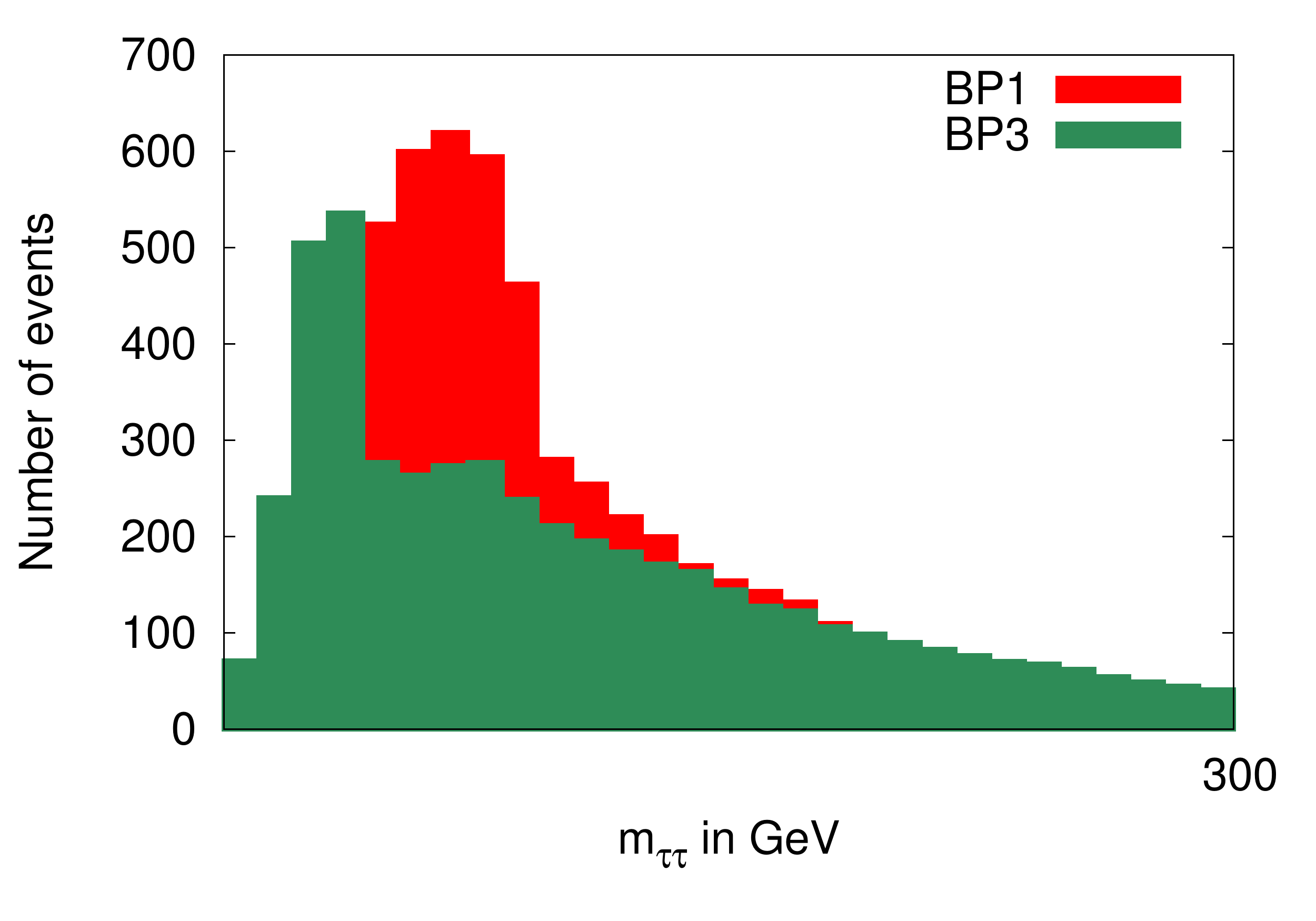} 	%\hspace*{-1cm}
		\includegraphics[width=0.32\linewidth, angle=-90]{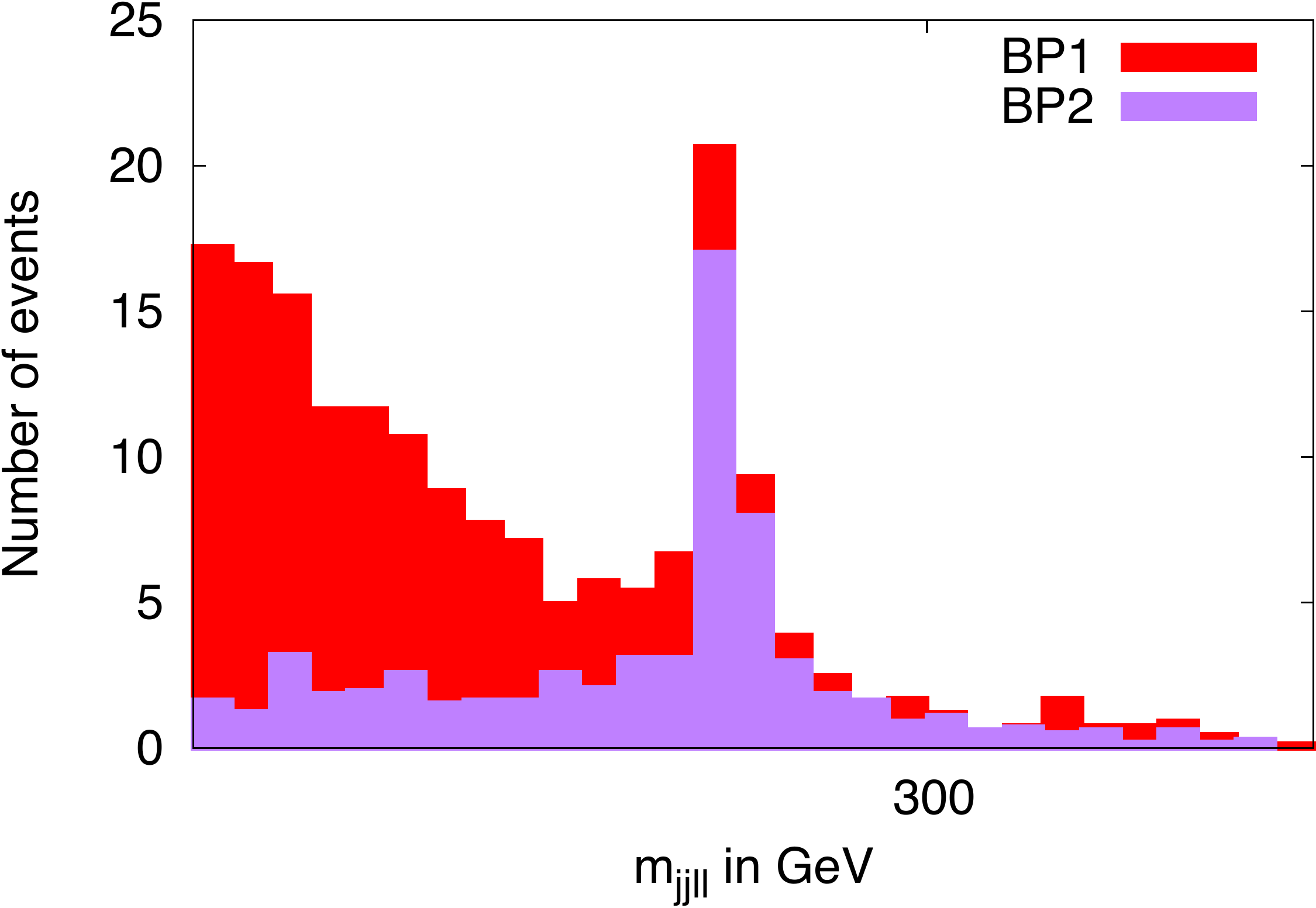}
		\caption{ $\tau\tau$ invariant mass distribution (left panel) and $jj\ell\ell$ invariant mass distribution (right panel) for the benchmark points at an integrated luminosity of 100\,fb$^{-1}$ at the LHC with 14\,TeV center of mass energy.}\label{disinv}
	\end{center}
\end{figure}

%%%%%%%%%%%%%%%%%%%%%%%%%%%%%%%%%%%%%%%%%%%%%%%%%%%%%%%%%%%%

In this section we probe the $AH^\pm$ production mode which follows the following decay chain leading to $2\tau + 2j +2 \ell$ final state.
$$A H^\pm  \to \tau^+ \tau^- N_i \ell^\pm \to 2 \tau + \ell^\pm W^\pm \ell^\pm \to 2 \tau + 2j +2\ell$$
We reconstruct the light pseudoscalar with $m_{\tau_{\rm{jet}}, \tau_{\rm{jet}}}$ invariant mass from hadronically reconstructed $\tau$ jets. Figure~\ref{disinv} (left panel) shows the invariant pseudoscalar mass for BP1.
Demanding $|m_{jj}-m_W|\leq 10$\,GeV i.e., the di-jet coming from $W^\pm$ boson, we can construct the $W^\pm$ and the pseudoscalar $A$ separately. As a next step, we select the events with di-jets from that window and the lepton to construct invariant mass $m_{jj\ell}$.  Then we look for the peak of the RHN $N_i$ in the  invariant mass distribution of $m_{jj\ell}$. Once we get the RHN mass peak,  we then construct $m_{jj \ell^+\ell^-}$, selecting events within 15\,GeV of  the peak of RHN with the remaining lepton, supposedly coming from the charged Higgs decay. The distribution for BP1 and BP2 are given in Fig.~\ref{disinv} (right panel) at 100\,fb$^{-1}$ of integrated luminosity at the LHC with 14\,TeV center of mass energy. It is clearly seen that both of the invariant mass are quite visible at $250$\,GeV. The $\pm 10$ GeV window near the peak consists of 30 and 25 events for BP1 and BP2, respectively. Interestingly the invariant mass distribution with the demand of $2\tau +2\ell$ plus the additional cuts is background free. Thus such points can reconstruct the charged Higgs mass peak with $\lesssim 1000$\,fb$^{-1}$ integrated luminosity. 

For BP3 the major decay modes for the charged Higgs bosons are into $AW^\pm$ and $N_i\ell_j$ but in this case the RHN decays into $A \nu_i$. We loose some amount of momentum as missing energy. Furthermore we lose more momentum as missing momentum from tau decays. This spoils the reconstruction of the RHN mass peak and so of the charged Higgs boson via $m_{\tau\tau~\ptmiss\ell}$.  Nevertheless, the information of the light pseudoscalar from $m_{\tau\tau}$ invariant mass can easily be probed here as well.

\section{Conclusions}
In this article we probe an additional decay channel of the charged Higgs boson decaying into a RHN  and a charged lepton. Such non-standard decay mode changes the current lower bound of the charged Higgs boson mass. To be explicit, we have considered Type-X 2HDM, where  a light pseudoscalar Higgs boson is still allowed, which opens up additional decay modes of  charged Higgs boson to $A W^\pm$ and RHN to $A \nu$ states. For relatively heavy pseudoscalar mass we have considered di-tau plus tri-lepton final states with different lepton flavour combination. We have shown from a PYTHIA based signal background analysis that $\gtrapprox5\sigma$ significance can be achieved for all four benchmark points at an integrated luminosity of 100\,fb$^{-1}$. For di-tau plus di-lepton signal, such significance can be achieved with very early data at the LHC with 14\,TeV center of mass energy. It is interesting to note that the inverse seesaw Yukawa coupling can be probed down to $Y_N \sim 0.2$, within 15\% systematic uncertainty, at HL LHC with 3000\,fb$^{-1}$ integrated luminosity for this channel.  We find that tagging four taus with one lepton (muon or electron) can also reach $5\sigma$ signal significance for all the benchmark points except BP2.  However, the results for $4\tau+ 2\ell$ does not look that promising for any of the benchmark points. Finally we leave it to the experimentalist to calculate the data driven QCD backgrounds, which may contribute via mis-tagging  of QCD jets and the subsequent  refinement of signal significance,  as this is beyond the scope of this analysis.

Next we focus on reconstructing the di-tau invariant mass as shown in Fig.~\ref{disinv} (left panel). It is evident from the figure that both light and heavy pseudoscalar masses can be reconstructed (BP3 and BP1) here. Followed by that we reconstruct the charged Higgs boson from the decay mode of charged Higgs boson to a RHN plus a charged lepton. We see for BP1 and BP2 it is quite possible to reconstruct the charged Higgs boson mass, whereas for BP3 due to large number of missing momentum, viz. neutrinos arising from the decays of RHN and taus, it is not possible to reconstruct such mass peak.  

This article thus provides a novel aspect of the charged Higgs boson decaying to RHNs plus a charged lepton. This non-standard decay mode of the charged Higgs boson can be introduced in other types of 2HDM and supersymmetric models. One can thus use these search strategies to test the respective scenarios. 
\label{sec:conclusion}

\section*{Acknowledgment} \vspace*{-0.3cm}
PB acknowledges IMSc, Chennai and KIAS, Seoul for the visits which were crucial in finishing the project and also SERB CORE Grant CRG/2018/004971. The work of RM has been supported in part by Grants No. FPA2014-53631-C2-1-P, FPA2017-84445-P and SEV-2014-0398 (AEI/ERDF, EU) and by PROMETEO/2017/053 (GV, ES).

%%%%%%%%%%%%%%%%%%%%%%%%%%%%%%%%%

\end{document}